# The Agda Universal Algebra Library
# Part 2: Structure

**Homomorphisms, terms, classes of algebras, subalgebras, and homomorphic images**

**William DeMeo**
Department of Algebra, Charles University in Prague

── **Abstract** ──────────────────────────────────────────
The Agda Universal Algebra Library (UALib) is a library of types and programs (theorems and proofs) we developed to formalize the foundations of universal algebra in dependent type theory using the Agda programming language and proof assistant. The UALib includes a substantial collection of definitions, theorems, and proofs from universal algebra, equational logic, and model theory, and as such provides many examples that exhibit the power of inductive and dependent types for representing and reasoning about mathematical structures and equational theories. In this paper, we describe the the types and proofs of the UALib that concern homomorphisms, terms, and subalgebras.

## Contents

# 1    Introduction

To support formalization in type theory of research level mathematics in universal algebra and related fields, we present the Agda Universal Algebra Library (AgdaUALib), a software library containing formal statements and proofs of the core definitions and results of universal algebra. The UALib is written in Agda [8], a programming language and proof assistant based on Martin-Löf Type Theory that not only supports dependent and inductive types, as well as proof tactics for proving things about the objects that inhabit these types.

## 1.1   Motivation

The seminal idea for the AgdaUALib project was the observation that, on the one hand, a number of fundamental constructions in universal algebra can be defined recursively, and theorems about them proved by structural induction, while, on the other hand, inductive and dependent types make possible very precise formal representations of recursively defined objects, which often admit elegant constructive proofs of properties of such objects. An important feature of such proofs in type theory is that they are total functional programs and, as such, they are computable, composable, and machine-verifiable.

Finally, our own research experience has taught us that a proof assistant and programming language (like Agda), when equipped with specialized libraries and domain-specific tactics to automate the proof idioms of our field, can be an extremely powerful and effective asset. As such we believe that proof assistants and their supporting libraries will eventually become indispensable tools in the working mathematician's toolkit.

## 1.2   Attributions and Contributions

The mathematical results described in this paper have well known *informal* proofs. Our main contribution is the formalization, mechanization, and verification of the statements and proofs of these results in dependent type theory using Agda.

Unless explicitly stated otherwise, the Agda source code described in this paper is due to the author, with the following caveat: the UALib depends on the Type Topology library of Martín Escardó [5]. Each dependency is carefully accounted for and mentioned in this paper. For the sake of completeness and clarity, and to keep the paper mostly self-contained, we repeat some definitions from the Type Topology library, but in each instance we cite the original source.[1]

## 1.3   Organization of the paper

In this paper we limit ourselves to the presentation of the middle third of the UALib, which includes types for representing homomorphisms, terms, and subalgebras. This limitation will allow us the space required to discuss some of the more interesting type theoretic and foundational issues that arise when developing a library of this kind and when attempting to represent advanced mathematical notions in type theory and formalize them in Agda.

This is the second installment of a three-part series of papers describing the AgdaUALib. The first part is [3] which covers relations, algebras, congruences, and quotients. The third part will cover free algebras, equational classes of algebras (i.e., varieties), and Birkhoff's HSP theorem.

---

[1]  In the UALib, such instances occur only inside hidden modules that are never actually used, followed immediately by a statement that imports the code in question from its original source.

The present paper is divided into three main parts (§2, §3, §4). The first of these introduces types representing *homomorphisms* from one algebra to another, and presents a formal statement and proof of the first fundamental theorem about homomorphisms, known as the *First Isomorphism Theorem*, as well as a version of the so-called *Second Isomorphism Theorem*. This is followed by dependent type definitions for representing *isomorphisms* and *homomorphic images* of algebraic structures.

In Section 3 we define inductive types to represent *terms* and the *term algebra* in a given signature. We prove the *universal property* of the term algebra which is the fact that term algebra is *free* (or *initial*) in the class of all algebras in the given signature. We define types that denote the interpretation of a term in an algebra type, called a *term operation*, including the interpretation of terms in *arbitrary products* of algebras (§3.2.1). We conclude § 3 with a subsection on the compatibility of terms with basic operations and congruence relations.

Section 4 presents inductive and dependent types for representing subuniverses and subalgebras of a given algebra. Here we define an inductive type that represents the *subuniverse generated by X*, for a given predicate $X$ : Pred | **A** | _,[2] and we use this type to formalize a few basic subuniverse lemmas. We also define types that pertain to arbitrary classes of algebras. In particular, in Subsection 4.2 on subalgebras, we define a type that represents the assertion that a given algebra is a subalgebra of some member of a class of algebras.

## 1.4 Resources

We conclude this introduction with some pointers to helpful reference materials. For the required background in Universal Algebra, we recommend the textbook by Clifford Bergman [1]. For the type theory background, we recommend the HoTT Book [9] and Escardó's Introduction to Univalent Foundations of Mathematics with Agda [5].

The following are informed the development of the UALib and are highly recommended.

- *Introduction to Univalent Foundations of Mathematics with Agda*, Escardó [5]
- *Dependent Types at Work*, Bove and Dybjer [2]
- *Dependently Typed Programming in Agda*, Norell and Chapman [7]
- *Formalization of Universal Algebra in Agda*, Gunther, Gadea, Pagano [6]
- *Programming Languages Foundations in Agda* [13]

More information about AgdaUALib can be obtained from the following official sources.

- `ualib.org` (the web site) documents every line of code in the library.
- `gitlab.com/ualib/ualib.gitlab.io` (the source code) AgdaUALib is open source.[3]
- *The Agda UALib, Part 1: equality, extensionality, truncation, and dependent types for relations and algebras* [3].
- *The Agda UALib, Part 3: free algebras, equational classes, and Birkhoff's theorem* [4].

The first item links to the official UALib html documentation which includes complete proofs of every theorem we mention here, and much more, including the Agda modules covered in the first and third installments of this series of papers on the UALib.

Finally, readers will get much more out of the paper if they download AgdaUALib from `https://gitlab.com/ualib/ualib.gitlab.io`, install the library, and try it out for themselves.

---

[2] As we learned in [3], such *X* represents a subset of the domain of the algebra **A**.

[3] License: Creative Commons Attribution-ShareAlike 4.0 International License.

## 2   Homomorphism Types

In this section we define types that represent some of the most important concepts from general (universal) algebra. Note that the Agda modules we describe here, and in succeeding sections depend on and import the modules that were presented in Part 1 of this series of three papers describing the AgdaUALib (see [3]).

We begin in Subsection 2.1 with the basic definition of *homomorphism*. In §2.2 we formalize the statement and proof of the first fundamental theorem about homomorphisms, which is sometimes referred to as the *First Isomorphism Theorem*. This is followed by §2.3, in which we define the type of *isomorphisms* between algebraic structures. Finally, in §2.4, we define types that manifest the notion of *homomorphic image*.

### 2.1   Basic definitions

This section presents the Homomorphisms.Basic module of the AgdaUALib, slightly abridged.[4] Since this is the first module we introduce in the installment of our series of papers documenting the AgdaUALib, we will begin by showing the start of the module in full. In later modules, we will leave such details implicit. Here is how the file `Homomorphisms/Basic.lagda` of the UALib begins.

```
{-# OPTIONS --without-K --exact-split --safe #-}

open import Algebras.Signatures using (Signature; 𝓞; 𝓥)
open import MGS-Subsingleton-Theorems using (global-dfunext)

module Homomorphisms.Basic {S : Signature 𝓞 𝓥}{gfe : global-dfunext} where

open import Algebras.Congruences{S = S} public
open import MGS-MLTT using (_≡⟨_⟩_; _∎) public
```

The OPTIONS *pragma* sets some parameters that specify the type theoretic foundations that we assume. These are discussed in [3, §2.1], but let's briefly review: *–without-K* disables Streicher's K axiom (see [10]); *–exact-split* makes Agda accept only definitions that use *definitional* equalities (see [12]); *–safe* ensures that nothing is postulated outright, so that every non-MLTT axiom has to be an explicit assumption (see [11] and [12]).[5]

#### 2.1.1   Homomorphisms

If **A** and **B** are *S*-algebras, then a *homomorphism* is a function $h : | \mathbf{A} | \to | \mathbf{B} |$ from the domain of **A** to the domain of **B** that is *compatible* (or *commutes*) with all of the basic operations of the signature; that is, for all operation symbols $f : | S |$ and all tuples $a : \| S \| f \to | \mathbf{A} |$, the following equality holds:[6]

$$h ((f \ \hat{}\  \mathbf{A})\ a) \equiv (f \ \hat{}\  \mathbf{B})\ (h \circ a).$$

Instead of "homomorphism," we sometimes use the nickname "hom" to refer to such a map.

---

[4]  For unabridged docs see `https://ualib.gitlab.io/Homomorphisms.Basic.html`; for source code see `https://gitlab.com/ualib/ualib.gitlab.io/-/blob/master/UALib/Homomorphisms/Basic.lagda`.
[5]  As in [3], MLTT stands for Martin-Löf Type Theory.
[6]  Recall, $h \circ a$ is the tuple whose *i*-th component is $h\ (a\ i)$.

To formalize the notion of homomorphism in type theory, we first define a type representing the assertion that a function $h : \mid \mathbf{A} \mid \to \mid \mathbf{B} \mid$, from the domain of $\mathbf{A}$ to the domain of $\mathbf{B}$, *commutes* (or is *compatible*) with an operation $f$, interpreted in the algebras $\mathbf{A}$ and $\mathbf{B}$. Pleasingly, the defining equation of the previous paragraph can be expressed in Agda unadulterated.[7]

```
module _ {𝒰 𝒲 : Universe} where

  compatible-op-map : (𝐀 : Algebra 𝒰 S)(𝐁 : Algebra 𝒲 S)
                      (f : ∣ S ∣)(h : ∣ 𝐀 ∣ → ∣ 𝐁 ∣) → 𝒱 ⊔ 𝒰 ⊔ 𝒲 ̇
  compatible-op-map 𝐀 𝐁 f h = ∀ a → h ((f ̂ 𝐀) a) ≡ (f ̂ 𝐁) (h ∘ a)
```

Note the appearance of the shorthand $\forall\, a$ in the definition of compatible-op-map. We can get away with this in place of $(a : \parallel S \parallel f \to \mid \mathbf{A} \mid)$ since Agda is able to infer that the $a$ here must be a tuple on $\mid \mathbf{A} \mid$ of "length" $\parallel S \parallel f$ (the arity of $f$).

We now define the type hom $\mathbf{A}\ \mathbf{B}$ of homomorphisms from $\mathbf{A}$ to $\mathbf{B}$ by first defining the property is-homomorphism.

```
  is-homomorphism : (𝐀 : Algebra 𝒰 S)(𝐁 : Algebra 𝒲 S) → (∣ 𝐀 ∣ → ∣ 𝐁 ∣) → 𝒪 ⊔ 𝒱 ⊔ 𝒰 ⊔ 𝒲 ̇
  is-homomorphism 𝐀 𝐁 g = ∀ f → compatible-op-map 𝐀 𝐁 f g

  hom : Algebra 𝒰 S → Algebra 𝒲 S → 𝒪 ⊔ 𝒱 ⊔ 𝒰 ⊔ 𝒲 ̇
  hom 𝐀 𝐁 = Σ g : (∣ 𝐀 ∣ → ∣ 𝐁 ∣) , is-homomorphism 𝐀 𝐁 g
```

### 2.1.2 Examples

Here we give a few very special examples of homomorphisms. In each case, the function in question commutes with the basic operations of *all* algebras and so, no matter the algebras involved, is always a homomorphism (trivially).

The most obvious example of a homomorphism is the identity map, which is proved to be a homomorphism as follows.

```
  𝒾𝒹 : {𝒰 : Universe} (A : Algebra 𝒰 S) → hom A A
  𝒾𝒹 _ = (λ x → x) , λ _ _ → 𝓇ℯ𝒻ℓ

  id-is-hom : {𝒰 : Universe}{𝐀 : Algebra 𝒰 S} → is-homomorphism 𝐀 𝐀 (𝒾𝒹 ∣ 𝐀 ∣)
  id-is-hom = λ _ _ → 𝓇ℯ𝒻ℓ
```

Next, lift and lower, defined in the Prelude.Lifts module, are (the maps of) homomorphisms. Again, the proof is trivial in each case.

```
  lift-is-hom : {𝐀 : Algebra 𝒰 S}{𝒲 : Universe} → is-homomorphism 𝐀 (lift-alg 𝐀 𝒲) lift
  lift-is-hom _ _ = 𝓇ℯ𝒻ℓ

  ℓ𝒾𝒻𝓉 : {𝐀 : Algebra 𝒰 S}{𝒲 : Universe} → hom 𝐀 (lift-alg 𝐀 𝒲)
  ℓ𝒾𝒻𝓉 = (lift , lift-is-hom)
```

---

[7] Here we put the definition inside an *anonymous module*, which starts with the module keyword followed by an underscore (instead of a module name). The purpose is simply to move the postulated typing judgments (the "parameters" of the module, e.g., 𝒰 𝒲 : Universe) out of the way so they don't obfuscate the definitions inside the module. In descriptions of the UALib, such as the present paper, we usually don't show the module declarations unless we wish to emphasize the typing judgments that are postulated in the module declaration.

```
lower-is-hom : {A : Algebra 𝒰 S}{𝒲 : Universe} → is-homomorphism (lift-alg A 𝒲) A lower
lower-is-hom _ _ = refl

𝓁𝓸𝓌𝓮𝓻 : (A : Algebra 𝒰 S){𝒲 : Universe} → hom (lift-alg A 𝒲) A
𝓁𝓸𝓌𝓮𝓻 A = (lower , lower-is-hom{A})
```

### 2.1.3   Monomorphisms and epimorphisms

We represent *monomorphisms* (injective homomorphisms) and *epimorphisms* (surjective homomorphisms) by the following types.

```
is-monomorphism : (A : Algebra 𝒰 S)(B : Algebra 𝒲 S) → (| A | → | B |) → 𝒪 ⊔ 𝒱 ⊔ 𝒰 ⊔ 𝒲 ̇
is-monomorphism A B g = is-homomorphism A B g × Monic g

mon : Algebra 𝒰 S → Algebra 𝒲 S → 𝒪 ⊔ 𝒱 ⊔ 𝒰 ⊔ 𝒲 ̇
mon A B = Σ g : (| A | → | B |) , is-monomorphism A B g

is-epimorphism : (A : Algebra 𝒰 S)(B : Algebra 𝒲 S) → (| A | → | B |) → 𝒪 ⊔ 𝒱 ⊔ 𝒰 ⊔ 𝒲 ̇
is-epimorphism A B g = is-homomorphism A B g × Epic g

epi : Algebra 𝒰 S → Algebra 𝒲 S  → 𝒪 ⊔ 𝒱 ⊔ 𝒰 ⊔ 𝒲 ̇
epi A B = Σ g : (| A | → | B |) , is-epimorphism A B g
```

It will be convenient to have a function that takes an inhabitant of 'mon' (or 'epi') and extracts the homomorphism part, or *hom reduct*, that is, the pair consisting of the map and a proof that the map is a homomorphism.

```
mon-to-hom : (A : Algebra 𝒰 S){B : Algebra 𝒲 S} → mon A B → hom A B
mon-to-hom A ϕ = | ϕ | , fst ‖ ϕ ‖

epi-to-hom : {A : Algebra 𝒰 S}(B : Algebra 𝒲 S) → epi A B → hom A B
epi-to-hom _ ϕ = | ϕ | , fst ‖ ϕ ‖
```

### 2.1.4   Equalizers in Agda

Recall, the equalizer of two functions (resp., homomorphisms) $g\ h : A \to B$ is the subset of $A$ on which the values of the functions $g$ and $h$ agree. We define the equalizer of functions and homomorphisms in the UALib as follows.

```
𝐸 : {B : Algebra 𝒲 S}(g h : | A | → | B |) → Pred | A | 𝒲
𝐸 g h x = g x ≡ h x

𝐸hom : (B : Algebra 𝒲 S)(g h : hom A B) → Pred | A | 𝒲
𝐸hom _ g h x = | g | x ≡ | h | x
```

We will define subuniverses in the Subalgebras.Subuniverses module, but we note here that the equalizer of homomorphisms from **A** to **B** will turn out to be subuniverse of **A**. Indeed, this is easily proved as follows.

```
𝐸hom-closed : (B : Algebra 𝒲 S)(g h : hom A B)
    →           ∀ f a → (∀ x → a x ∈ 𝐸hom B g h)
                ———————————————————————————
    →           | g | ((f ̂ A) a) ≡ | h | ((f ̂ A) a)
```

```
Ehom-closed B g h f a p = | g | ((f ˆ A) a)      ≡⟨ ∥ g ∥ f a ⟩
                          (f ˆ B)(| g | ∘ a)     ≡⟨ ap (f ˆ B)(gfe p) ⟩
                          (f ˆ B)(| h | ∘ a)     ≡⟨ (∥ h ∥ f a)⁻¹ ⟩
                          | h | ((f ˆ A) a)      ∎
```

### 2.1.5  Kernels of Homomorphisms

The *kernel* of a homomorphism is a congruence relation and conversely for every congruence relation $\theta$, there exists a homomorphism with kernel $\theta$ (namely, that canonical projection onto the quotient modulo $\theta$).

```
homker-compatible : {A : Algebra 𝒰 S}(B : Algebra 𝒲 S)(h : hom A B)
  →                compatible A (KER-rel | h |)

homker-compatible {A} B h f {u}{v} Kerhab = γ
  where
  γ : | h | ((f ˆ A) u) ≡ | h | ((f ˆ A) v)
  γ = | h | ((f ˆ A) u)    ≡⟨ ∥ h ∥ f u ⟩
      (f ˆ B)(| h | ∘ u)   ≡⟨ ap (f ˆ B)(gfe λ x → Kerhab x) ⟩
      (f ˆ B)(| h | ∘ v)   ≡⟨ (∥ h ∥ f v)⁻¹ ⟩
      | h | ((f ˆ A) v)    ∎

homker-equivalence : {A : Algebra 𝒰 S}(B : Algebra 𝒲 S)(h : hom A B)
  →                 IsEquivalence (KER-rel | h |)

homker-equivalence A h = map-kernel-IsEquivalence | h |
```

It is convenient to define a function that takes a homomorphism and constructs a congruence from its kernel. We call this function kercon.

```
kercon : {A : Algebra 𝒰 S}(B : Algebra 𝒲 S)(h : hom A B) → Congruence A
kercon B h = mkcon (KER-rel | h |)(homker-compatible B h)(homker-equivalence B h)
```

With this congruence we define the corresponding quotient, along with some syntactic sugar to denote it.

```
kerquo : {A : Algebra 𝒰 S}(B : Algebra 𝒲 S)(h : hom A B) → Algebra (𝒰 ⊔ 𝒲 ⁺) S
kerquo {A} B h = A ╱ (kercon B h)

_[_]/ker_ : (A : Algebra 𝒰 S)(B : Algebra 𝒲 S)(h : hom A B) → Algebra (𝒰 ⊔ 𝒲 ⁺) S
A [ B ]/ker h = kerquo {A} B h
```

Thus, given $h :$ hom A B, we can construct the quotient of A modulo the kernel of $h$, and the UALib syntax for this quotient is A [ B ]/ker $h$.

### 2.1.6  The natural projection

Given an algebra A and a congruence $\theta$, the *natural* or *canonical projection* is a map from A onto A ╱ $\theta$ that is constructed, and proved epimorphic, as follows.

```
πepi : {A : Algebra 𝒰 S} (θ : Congruence{𝒰}{𝒲} A) → epi A (A ╱ θ)
πepi {A} θ = cπ , cπ-is-hom , cπ-is-epic where

  cπ : | A | → | A ╱ θ |
```

```
cπ a = [[ a ]]{⟨ θ ⟩}

cπ-is-hom : is-homomorphism A (A ╱ θ) cπ
cπ-is-hom _ _ = refl

cπ-is-epic : Epic cπ
cπ-is-epic (.(⟨ θ ⟩ a) , a , refl) = Image_∋_.im a
```

Sometimes we don't care about the surjectivity of πepi and we want to work with its *homomorphic reduct*, which is obtained by applying epi-to-hom, like so.

```
πhom : {A : Algebra 𝒰 S}(θ : Congruence{𝒰}{𝒲} A) → hom A (A ╱ θ)
πhom {A} θ = epi-to-hom (A ╱ θ) (πepi θ)
```

We combine the foregoing to define a function that takes $S$-algebras **A** and **B**, and a homomorphism $h$ : hom **A** **B** and returns the canonical epimorphism from **A** onto the quotient algebra **A** [ **B** ]/ ker $h$ (which, recall, is **A** modulo the kernel of $h$).

```
πker : {A : Algebra 𝒰 S}(B : Algebra 𝒲 S)(h : hom A B) → epi A (A [ B ]/ker h)
πker {A} B h = πepi (kercon B h)
```

The kernel of the canonical projection of **A** onto **A** ╱ $θ$ is equal to $θ$, but since equality of inhabitants of certain types (like Congruence or Rel) can be a tricky business, we settle for proving the containment **A** ╱ $θ$ ⊆ $θ$. Of the two containments, this is the easier one to prove; luckily it is also the one we need later.

```
ker-in-con : (A : Algebra 𝒰 S)(θ : Congruence{𝒰}{𝒲} A)(x y : | A |)
    →       ⟨ kercon (A ╱ θ) (πhom θ) ⟩ x y → ⟨ θ ⟩ x y

ker-in-con A θ x y hyp = ╱-refl θ hyp
```

### 2.1.7   Product homomorphisms

Suppose we have an algebra **A** : Algebra 𝒰 $S$, a type $I$ : 𝒥 ·, and a family ℬ : $I$ → Algebra 𝒲 $S$ of algebras. We sometimes refer to the inhabitants of $I$ as *indices*, and call ℬ an *indexed family of algebras*. If in addition we have a family ℏ : $(i : I)$ → hom **A** (ℬ $i$) of homomorphisms, then we can construct a homomorphism from **A** to the product ⊓ ℬ in the natural way.

```
⊓-hom-co : {A : Algebra 𝒰 S}{I : 𝒥 ·}(ℬ : I → Algebra 𝒲 S)
    →       Π i : I , hom A (ℬ i) → hom A (⊓ ℬ)

⊓-hom-co {A} ℬ ℏ = ϕ , ϕhom
    where
    ϕ : | A | → | ⊓ ℬ |
    ϕ a = λ i → | ℏ i | a

    ϕhom : is-homomorphism A (⊓ ℬ) ϕ
    ϕhom f 𝒶 = gfe λ i → ∥ ℏ i ∥ f 𝒶
```

The family ℏ of homomorphisms inhabits the dependent type Π $i : I$, hom **A** (ℬ $i$). The syntax we use to represent this type is available to us because of the way -Π is defined in the Type Topology library. We like this syntax because it is very close to the notation one finds in the standard type theory literature. However, we could equally well have used one of the following alternatives, which may be closer to "standard Agda" syntax:

    Π λ i → hom **A** (𝓑 i)   or   (i : I) → hom **A** (𝓑 i)   or   ∀ i → hom **A** (𝓑 i).

The foregoing generalizes easily to the case in which the domain is also a product of a family of algebras. That is, if we are given 𝓐 : I → Algebra 𝓤 S and 𝓑 : I → Algebra 𝓦 S (two families of S-algebras), and ℏ : Π i : I , hom (𝓐 i) (𝓑 i) (a family of homomorphisms), then we can construct a homomorphism from ∏ 𝓐 to ∏ 𝓑 in the following natural way.

```
∏-hom : {I : 𝓘 ˙}(𝓐 : I → Algebra 𝓤 S)(𝓑 : I → Algebra 𝓦 S)
    →       Π i : I , hom (𝓐 i)(𝓑 i) → hom (∏ 𝓐)(∏ 𝓑)

∏-hom 𝓐 𝓑 ℏ = φ , φhom
 where
 φ : | ∏ 𝓐 | → | ∏ 𝓑 |
 φ = λ x i → | ℏ i | (x i)

 φhom : is-homomorphism (∏ 𝓐) (∏ 𝓑) φ
 φhom f 𝒶 = gfe λ i → ∥ ℏ i ∥ f (λ x → 𝒶 x i)
```

### 2.1.8 Projection homomorphisms

Later we will need a proof of the fact that the natural projection out of a product algebra onto one of its factors is a homomorphism.

```
∏-projection-hom : {I : 𝓘 ˙}(𝓑 : I → Algebra 𝓦 S) → Π i : I , hom (∏ 𝓑) (𝓑 i)

∏-projection-hom 𝓑 = λ i → ℏ i , ℏhom i
 where
 ℏ : ∀ i → | ∏ 𝓑 | → | 𝓑 i |
 ℏ i = λ x → x i

 ℏhom : ∀ i → is-homomorphism (∏ 𝓑) (𝓑 i) (ℏ i)
 ℏhom _ _ _ = 𝓇ℯ𝒻ℓ
```

Of course, we could prove a more general result involving projections onto multiple factors, but so far the single-factor result has sufficed.

## 2.2 Homomorphism Theorems

This section presents the Homomorphisms.Noether module of the AgdaUALib, slightly abridged.[8]

### 2.2.1 The First Homomorphism Theorem

Here is a version of the so-called *First Homomorphism theorem*, sometimes called Noether's First Homomorphism theorem, after Emmy Noether who was among the first proponents of the abstract approach to algebra that we now refer to as "modern algebra."

```
open Congruence
module _ {𝓤 𝓦 : Universe}
```

---

[8] For unabridged docs see `https://ualib.gitlab.io/Homomorphisms.Noether.html`; for source code see `https://gitlab.com/ualib/ualib.gitlab.io/-/blob/master/UALib/Homomorphisms/Noether.lagda`.

        *– extensionality assumptions –*
        ($fe$ : dfunext $\mathcal{V}$ $\mathcal{W}$) ($pe$ : prop-ext $\mathcal{U}$ $\mathcal{W}$)

        (**A** : Algebra $\mathcal{U}$ $S$)(**B** : Algebra $\mathcal{W}$ $S$)($h$ : hom **A** **B**)

        *– truncation assumptions –*
        ($Bset$ : is-set ∣ **B** ∣)
        ($ssR$ : ∀ $a$ $x$ → is-subsingleton (⟨ kercon **B** $h$ ⟩ $a$ $x$))
        ($ssA$ : ∀ $C$ → is-subsingleton ($\mathscr{C}${$A$ = ∣ **A** ∣}{⟨ kercon **B** $h$ ⟩} $C$))

where

FirstHomomorphismTheorem :

  Σ $\phi$ : hom (**A** [ **B** ]/ker $h$) **B** ,
      (∣ $h$ ∣ ≡ ∣ $\phi$ ∣ ∘ ∣ $\pi$ker **B** $h$ ∣) × Monic ∣ $\phi$ ∣ × is-embedding ∣ $\phi$ ∣

FirstHomomorphismTheorem = ($\phi$ , $\phi$hom) , $\phi$com , $\phi$mon , $\phi$emb
  where
  $\theta$ : Congruence **A**
  $\theta$ = kercon **B** $h$

  $\phi$ : ∣ **A** [ **B** ]/ker $h$ ∣ → ∣ **B** ∣
  $\phi$ $a$ = ∣ $h$ ∣ ⌜ $a$ ⌝

  **R** : Pred$_2$ ∣ **A** ∣ $\mathcal{W}$
  **R** = ⟨ kercon **B** $h$ ⟩ , $ssR$

  $\phi$hom : is-homomorphism (**A** [ **B** ]/ker $h$) **B** $\phi$
  $\phi$hom $f$ **a** = ∣ $h$ ∣ ( ( $f$ $\hat{\ }$ **A**) ($\lambda$ $x$ → ⌜ **a** $x$ ⌝ ) ) ≡⟨ ∥ $h$ ∥ $f$ ($\lambda$ $x$ → ⌜ **a** $x$ ⌝) ⟩
         ($f$ $\hat{\ }$ **B**) (∣ $h$ ∣ ∘ ($\lambda$ $x$ → ⌜ **a** $x$ ⌝)) ≡⟨ ap ($f$ $\hat{\ }$ **B**) ($fe$ $\lambda$ $x$ → $refl$) ⟩
         ($f$ $\hat{\ }$ **B**) ($\lambda$ $x$ → $\phi$ (**a** $x$)) ∎

  $\phi$mon : Monic $\phi$
  $\phi$mon (.(⟨ $\theta$ ⟩ $u$) , $u$ , refl _) (.(⟨ $\theta$ ⟩ $v$) , $v$ , refl _) $\phi uv$ =
    class-extensionality' {**R** = **R**} $pe$ $ssA$ (IsEquiv $\theta$) $\phi uv$

  $\phi$com : ∣ $h$ ∣ ≡ $\phi$ ∘ ∣ $\pi$ker **B** $h$ ∣
  $\phi$com = $refl$

  $\phi$emb : is-embedding $\phi$
  $\phi$emb = monic-is-embedding|sets $\phi$ $Bset$ $\phi$mon

Next we show that the homomorphism $\phi$, whose existence we just proved, is unique.

  NoetherHomUnique : ($f$ $g$ : hom (**A** [ **B** ]/ker $h$) **B**)
  →                    ∣ $h$ ∣ ≡ ∣ $f$ ∣ ∘ ∣ $\pi$ker **B** $h$ ∣ → ∣ $h$ ∣ ≡ ∣ $g$ ∣ ∘ ∣ $\pi$ker **B** $h$ ∣
  →                    ∀ $a$ → ∣ $f$ ∣ $a$ ≡ ∣ $g$ ∣ $a$

  NoetherHomUnique $f$ $g$ $hfk$ $hgk$ (.(⟨ kercon **B** $h$ ⟩ $a$) , $a$ , $refl$) =

    let $\theta$ = (⟨ kercon **B** $h$ ⟩ $a$ , $a$ , $refl$) in

    ∣ $f$ ∣ $\theta$ ≡⟨ cong-app ($hfk$ ⁻¹) $a$ ⟩ ∣ $h$ ∣ $a$ ≡⟨ cong-app ($hgk$) $a$ ⟩ ∣ $g$ ∣ $\theta$ ∎

If we postulate function extensionality, then we have[9]

```
fe-NoetherHomUnique : funext (𝒰 ⊔ 𝒲 ⁺) 𝒲 → (f g : hom (A [ B ]/ker h) B)
  →                   | h | ≡ | f | ∘ | πker B h | → | h | ≡ | g | ∘ | πker B h |
  →                   | f | ≡ | g |

fe-NoetherHomUnique fe f g hfk hgk = fe (NoetherHomUnique f g hfk hgk)
```

Assuming the hypotheses of the First Homomorphism theorem, if we add the assumption that $h$ is epic, then we get the so-called *First Isomorphism theorem*.

```
FirstIsomorphismTheorem :

  Epic | h | → Σ f : (epi (A [ B ]/ker h) B) , (| h | ≡ | f | ∘ | πker B h |) × is-embedding | f |

FirstIsomorphismTheorem hE = (fmap , fhom , fepic) , refl , femb
  where
  θ : Congruence A
  θ = kercon B h

  fmap : | A [ B ]/ker h | → | B |
  fmap ⟦a⟧ = | h | ⌜ ⟦a⟧ ⌝

  fhom : is-homomorphism (A [ B ]/ker h) B fmap
  fhom f a = | h |((f ̂ A) λ x → ⌜ a x ⌝) ≡⟨ ‖ h ‖ f (λ x → ⌜ a x ⌝) ⟩
             (f ̂ B)(| h | ∘ λ x → ⌜ a x ⌝) ≡⟨ ap(f ̂ B)(gfe λ _ → refl)⟩
             (f ̂ B) (fmap ∘ a) ∎

  fepic : Epic fmap
  fepic b = γ where
    a : | A |
    a = EpicInv | h | hE b

    bfa : b ≡ fmap ⟦ a ⟧
    bfa = (cong-app (EpicInvIsRightInv gfe | h | hE) b)⁻¹

    γ : Image fmap ∋ b
    γ = Image_∋_.eq b ⟦ a ⟧ bfa

  fmon : Monic fmap
  fmon (.(⟨ θ ⟩ a) , a , refl) (.(⟨ θ ⟩ a') , a' , refl) faa' =
    class-extensionality' {R = ⟨ kercon B h ⟩ , ssR} pe ssA (IsEquiv θ) faa'

  femb : is-embedding fmap
  femb = monic-is-embedding|sets fmap Bset fmon
```

The argument used above to prove NoetherHomUnique can also be used to prove uniqueness of the epimorphism $f$ found in the isomorphism theorem.

---

[9] We already assumed *global* function extensionality in this module, so we could just appeal to that in this case. However, we make a local function extensionality assumption explicit here merely to highlight where and how the principle is applied.

NoetherIsoUnique : $(f\ g : \mathsf{epi}\ (\mathbf{A}\ [\ \mathbf{B}\ ]/\mathsf{ker}\ h)\ \mathbf{B}) \to |\ h\ | \equiv |\ f\ | \circ |\ \pi\mathsf{ker}\ \mathbf{B}\ h\ |$
$\to\qquad |\ h\ | \equiv |\ g\ | \circ |\ \pi\mathsf{ker}\ \mathbf{B}\ h\ | \to \forall\ a \to |\ f\ |\ a \equiv |\ g\ |\ a$

NoetherIsoUnique $f\ g\ hfk\ hgk\ (.(\langle\ \mathsf{kercon}\ \mathbf{B}\ h\ \rangle\ a)\ ,\ a\ ,\ \mathit{refl}) =$

  let $\theta = (\langle\ \mathsf{kercon}\ \mathbf{B}\ h\ \rangle\ a\ ,\ a\ ,\ \mathit{refl})$ in

  $|\ f\ |\ \theta \equiv\langle\ \mathsf{cong\text{-}app}(hfk\ ^{-1})a\ \rangle\ |\ h\ |\ a \equiv\langle\ \mathsf{cong\text{-}app}(hgk)a\ \rangle\ |\ g\ |\ \theta\ \blacksquare$

### 2.2.2 Composition of homomorphisms

The composition of homomorphisms is again a homomorphism. There are a number of alternative ways to formalize this fact in Agda. The two representations included in the UALib are the following.

$\circ$-hom : $(\mathbf{A} : \mathsf{Algebra}\ \mathcal{X}\ S)\{\mathbf{B} : \mathsf{Algebra}\ \mathcal{Y}\ S\}(\mathbf{C} : \mathsf{Algebra}\ \mathcal{Z}\ S)$
$\to\qquad \mathsf{hom}\ \mathbf{A}\ \mathbf{B} \to \mathsf{hom}\ \mathbf{B}\ \mathbf{C} \to \mathsf{hom}\ \mathbf{A}\ \mathbf{C}$

$\circ$-hom $\mathbf{A}\ \{\mathbf{B}\}\ \mathbf{C}\ (g\ ,\ ghom)\ (h\ ,\ hhom) = h \circ g\ ,\ \gamma$ where

  $\gamma : \forall\ f\ a \to (h \circ g)((f\ \hat{}\ \mathbf{A})\ a) \equiv (f\ \hat{}\ \mathbf{C})(h \circ g \circ a)$

  $\gamma\ f\ a = (h \circ g)\ ((f\ \hat{}\ \mathbf{A})\ a) \equiv\langle\ \mathsf{ap}\ h\ (\ ghom\ f\ a\ )\ \rangle$
        $h\ ((f\ \hat{}\ \mathbf{B})\ (g \circ a)) \equiv\langle\ hhom\ f\ (\ g \circ a\ )\ \rangle$
        $(f\ \hat{}\ \mathbf{C})\ (h \circ g \circ a)\ \blacksquare$

$\circ$-is-hom : $(\mathbf{A} : \mathsf{Algebra}\ \mathcal{X}\ S)\{\mathbf{B} : \mathsf{Algebra}\ \mathcal{Y}\ S\}(\mathbf{C} : \mathsf{Algebra}\ \mathcal{Z}\ S)$
        $\{f : |\ \mathbf{A}\ | \to |\ \mathbf{B}\ |\}\ \{g : |\ \mathbf{B}\ | \to |\ \mathbf{C}\ |\}$
$\to\qquad \mathsf{is\text{-}homomorphism}\ \mathbf{A}\ \mathbf{B}\ f \to \mathsf{is\text{-}homomorphism}\ \mathbf{B}\ \mathbf{C}\ g$
$\to\qquad \mathsf{is\text{-}homomorphism}\ \mathbf{A}\ \mathbf{C}\ (g \circ f)$

$\circ$-is-hom $\mathbf{A}\ \mathbf{C}\ \{f\}\ \{g\}\ \mathit{fhom}\ \mathit{ghom} = \|\ \circ\text{-hom}\ \mathbf{A}\ \mathbf{C}\ (f\ ,\ \mathit{fhom})\ (g\ ,\ \mathit{ghom})\ \|$

### 2.2.3 Homomorphism decomposition

If $g : \mathsf{hom}\ \mathbf{A}\ \mathbf{B}$, $h : \mathsf{hom}\ \mathbf{A}\ \mathbf{C}$, $h$ is surjective, and $\mathsf{ker}\ h \subseteq \mathsf{ker}\ g$, then there exists $\phi : \mathsf{hom}\ \mathbf{C}\ \mathbf{B}$ such that $g = \phi \circ h$, that is, such that the following diagram commutes.

$$\begin{array}{ccc} \mathbf{A} & \xrightarrow{h} & \mathbf{C} \\ {}_{g}\searrow & & \swarrow_{\exists \phi} \\ & \mathbf{B} & \end{array}$$

This, or some variation of it, is sometimes referred to as the *Second Isomorphism Theorem*. We formalize its statement and proof as follows. (Notice that the proof is constructive.)

homFactor : $\{\mathcal{U} : \mathsf{Universe}\} \to \mathsf{funext}\ \mathcal{U}\ \mathcal{U} \to \{\mathbf{A}\ \mathbf{B}\ \mathbf{C} : \mathsf{Algebra}\ \mathcal{U}\ S\}$
        $(g : \mathsf{hom}\ \mathbf{A}\ \mathbf{B})\ (h : \mathsf{hom}\ \mathbf{A}\ \mathbf{C})$
$\to\qquad \mathsf{ker\text{-}pred}\ |\ h\ | \subseteq \mathsf{ker\text{-}pred}\ |\ g\ | \to \mathsf{Epic}\ |\ h\ |$
        ———————————————————-
$\to\qquad \Sigma\ \phi : (\mathsf{hom}\ \mathbf{C}\ \mathbf{B})\ ,\ |\ g\ | \equiv |\ \phi\ | \circ |\ h\ |$

```
homFactor fe{A}{B}{C}(g , ghom)(h , hhom) Kh⊆Kg hEpi = (ϕ , ϕIsHomCB) , g≡ϕ∘h
  where
  hInv : | C | → | A |
  hInv = λ c → (EpicInv h hEpi) c

  ϕ : | C | → | B |
  ϕ = λ c → g ( hInv c )

  ξ : ∀ x → ker-pred h (x , hInv (h x))
  ξ x = (cong-app (EpicInvIsRightInv fe h hEpi) (h x))⁻¹

  g≡ϕ∘h : g ≡ ϕ ∘ h
  g≡ϕ∘h = fe λ x → Kh⊆Kg (ξ x)

  ζ : (f : | S |)(𝒄 : ‖ S ‖ f → | C |)(x : ‖ S ‖ f) → 𝒄 x ≡ (h ∘ hInv)(𝒄 x)
  ζ f 𝒄 x = (cong-app (EpicInvIsRightInv fe h hEpi) (𝒄 x))⁻¹

  ι : (f : | S |)(𝒄 : ‖ S ‖ f → | C |) → 𝒄 ≡ h ∘ (hInv ∘ 𝒄)
  ι f 𝒄 = ap (λ - → - ∘ 𝒄)(EpicInvIsRightInv fe h hEpi)⁻¹

  useker : ∀ f 𝒄 → g(hInv (h((f ̂ A)(hInv ∘ 𝒄)))) ≡ g((f ̂ A)(hInv ∘ 𝒄))
  useker f c = Kh⊆Kg (cong-app(EpicInvIsRightInv fe h hEpi)(h ((f ̂ A)(hInv ∘ c))))

  ϕIsHomCB : (f : | S |)(𝒄 : ‖ S ‖ f → | C |) → ϕ((f ̂ C) 𝒄) ≡ (f ̂ B)(ϕ ∘ 𝒄)
  ϕIsHomCB f 𝒄 = g (hInv ((f ̂ C) 𝒄))                    ≡⟨ i  ⟩
                 g (hInv ((f ̂ C)(h ∘ (hInv ∘ 𝒄))))       ≡⟨ ii ⟩
                 g (hInv (h ((f ̂ A)(hInv ∘ 𝒄))))         ≡⟨ iii ⟩
                 g ((f ̂ A)(hInv ∘ 𝒄))                    ≡⟨ iv ⟩
                 (f ̂ B)(λ x → g (hInv (𝒄 x)))            ∎
    where
    i   = ap (g ∘ hInv) (ap (f ̂ C) (ι f 𝒄))
    ii  = ap (g ∘ hInv) (hhom f (hInv ∘ 𝒄))⁻¹
    iii = useker f 𝒄
    iv  = ghom f (hInv ∘ 𝒄)
```

Here's a more general version.

```
HomFactor : (A : Algebra 𝒳 S){B : Algebra 𝒴 S}{C : Algebra 𝒵 S}
            (β : hom A B) (γ : hom A C)
  →         Epic | γ | → (KER-pred | γ |) ⊆ (KER-pred | β |)
            ─────────────────────────────────────────────
  →         Σ ϕ : (hom C B) , | β | ≡ | ϕ | ∘ | γ |

HomFactor A {B}{C} β γ γE Kγβ = (ϕ , ϕIsHomCB) , βϕγ
  where
  γInv : | C | → | A |
  γInv = λ y → (EpicInv | γ | γE) y

  ϕ : | C | → | B |
  ϕ = λ y → | β | ( γInv y )
```

$\xi : (x : | \mathbf{A} |) \to \mathsf{KER\text{-}pred} | \gamma | (x, \gamma\mathsf{Inv} (| \gamma | x))$
$\xi\ x = (\mathsf{cong\text{-}app}\ (\mathsf{EpicInvIsRightInv}\ gfe | \gamma | \gamma E)\ (| \gamma | x\ ))^{-1}$

$\beta\phi\gamma : | \beta | \equiv \phi \circ | \gamma |$
$\beta\phi\gamma = gfe\ \lambda\ x \to K\gamma\beta\ (\xi\ x)$

$\iota : (f : | S |)(\mathbf{c} : \| S \| f \to | \mathbf{C} |) \to \mathbf{c} \equiv | \gamma | \circ (\gamma\mathsf{Inv} \circ \mathbf{c})$
$\iota\ f\ \mathbf{c} = \mathsf{ap}\ (\lambda\ \text{-} \to \text{-} \circ \mathbf{c})(\mathsf{EpicInvIsRightInv}\ gfe | \gamma | \gamma E)^{-1}$

$\mathsf{useker} : \forall\ f\ \mathbf{c} \to | \beta | (\gamma\mathsf{Inv} (| \gamma | ((f\ \hat{}\ \mathbf{A}) (\gamma\mathsf{Inv} \circ \mathbf{c})))) \equiv | \beta |((f\ \hat{}\ \mathbf{A}) (\gamma\mathsf{Inv} \circ \mathbf{c}))$
$\mathsf{useker}\ f\ \mathbf{c} = K\gamma\beta\ (\mathsf{cong\text{-}app}\ (\mathsf{EpicInvIsRightInv}\ gfe | \gamma | \gamma E)(| \gamma | ((f\ \hat{}\ \mathbf{A})(\gamma\mathsf{Inv} \circ \mathbf{c}))))$

$\phi\mathsf{IsHomCB} : \forall\ f\ \mathbf{c} \to \phi\ ((f\ \hat{}\ \mathbf{C})\ \mathbf{c}) \equiv ((f\ \hat{}\ \mathbf{B})(\phi \circ \mathbf{c}))$
$\phi\mathsf{IsHomCB}\ f\ \mathbf{c} = | \beta | (\gamma\mathsf{Inv} ((f\ \hat{}\ \mathbf{C})\ \mathbf{c}))$                                $\equiv\langle\ \text{i}\ \rangle$
$\quad | \beta | (\gamma\mathsf{Inv} ((f\ \hat{}\ \mathbf{C})(| \gamma | \circ (\gamma\mathsf{Inv} \circ \mathbf{c}))))$    $\equiv\langle\ \text{ii}\ \rangle$
$\quad | \beta | (\gamma\mathsf{Inv} (| \gamma | ((f\ \hat{}\ \mathbf{A})(\gamma\mathsf{Inv} \circ \mathbf{c}))))$    $\equiv\langle\ \text{iii}\ \rangle$
$\quad | \beta | ((f\ \hat{}\ \mathbf{A})(\gamma\mathsf{Inv} \circ \mathbf{c}))$                              $\equiv\langle\ \text{iv}\ \rangle$
$\quad ((f\ \hat{}\ \mathbf{B})(\lambda\ x \to | \beta | (\gamma\mathsf{Inv}\ (\mathbf{c}\ x))))$            ∎

where
i   $= \mathsf{ap}\ (| \beta | \circ \gamma\mathsf{Inv})\ (\mathsf{ap}\ (f\ \hat{}\ \mathbf{C})\ (\iota\ f\ \mathbf{c}))$
ii  $= \mathsf{ap}\ (| \beta | \circ \gamma\mathsf{Inv})\ (\| \gamma \| f\ (\gamma\mathsf{Inv} \circ \mathbf{c}))^{-1}$
iii $= \mathsf{useker}\ f\ \mathbf{c}$
iv  $= \| \beta \| f\ (\gamma\mathsf{Inv} \circ \mathbf{c})$

If, in addition, both $\beta$ and $\gamma$ are epic, then so is $\phi$.

$\mathsf{HomFactorEpi} : (\mathbf{A} : \mathsf{Algebra}\ \mathcal{X}\ S)\{\mathbf{B} : \mathsf{Algebra}\ \mathcal{Y}\ S\}\{\mathbf{C} : \mathsf{Algebra}\ \mathcal{Z}\ S\}$
$\qquad\qquad (\beta : \mathsf{hom}\ \mathbf{A}\ \mathbf{B})\ (\beta e : \mathsf{Epic} | \beta |)$
$\qquad\qquad (\xi : \mathsf{hom}\ \mathbf{A}\ \mathbf{C})\ (\xi e : \mathsf{Epic} | \xi |)$
$\to \qquad (\mathsf{KER\text{-}pred} | \xi |) \subseteq (\mathsf{KER\text{-}pred} | \beta |)$
$\to \qquad \Sigma\ \phi : (\mathsf{epi}\ \mathbf{C}\ \mathbf{B}) , | \beta | \equiv | \phi | \circ | \xi |$

$\mathsf{HomFactorEpi}\ \mathbf{A}\ \{\mathbf{B}\}\{\mathbf{C}\}\ \beta\ \beta e\ \xi\ \xi e\ kerincl = (\mathsf{fst} | \phi\mathsf{F} | , (\mathsf{snd} | \phi\mathsf{F} | , \phi\mathsf{E})) , \| \phi\mathsf{F} \|$
where
$\phi\mathsf{F} : \Sigma\ \phi : (\mathsf{hom}\ \mathbf{C}\ \mathbf{B}) , | \beta | \equiv | \phi | \circ | \xi |$
$\phi\mathsf{F} = \mathsf{HomFactor}\ \mathbf{A}\ \{\mathbf{B}\}\{\mathbf{C}\}\ \beta\ \xi\ \xi e\ kerincl$

$\xi\mathsf{inv} : | \mathbf{C} | \to | \mathbf{A} |$
$\xi\mathsf{inv} = \lambda\ c \to (\mathsf{EpicInv} | \xi | \xi e)\ c$

$\beta\mathsf{inv} : | \mathbf{B} | \to | \mathbf{A} |$
$\beta\mathsf{inv} = \lambda\ b \to (\mathsf{EpicInv} | \beta | \beta e)\ b$

$\phi : | \mathbf{C} | \to | \mathbf{B} |$
$\phi = \lambda\ c \to | \beta | (\xi\mathsf{inv}\ c)$

$\phi\mathsf{E} : \mathsf{Epic}\ \phi$
$\phi\mathsf{E} = \mathsf{epic\text{-}factor}\ gfe | \beta | | \xi | \phi \| \phi\mathsf{F} \| \beta e$

## 2.3 Isomorphisms

This section presents the Homomorphisms.Isomorphisms module of the AgdaUALib, slightly abridged.[10] Here we formalize the notion of *isomorphism* between algebraic structures.

### 2.3.1 Definition of isomorphism

Recall, $f \sim g$ means $f$ and $g$ are *extensionally* (or *point-wise*) *equal*; i.e., $\forall\ x,\ f\ x \equiv g\ x$. We use this notion of equality of functions in the following definition of *isomorphism*.

$\_\cong\_$ : $\{\mathcal{U}\ \mathcal{W}$ : Universe$\}$($\mathbf{A}$ : Algebra $\mathcal{U}\ S$)($\mathbf{B}$ : Algebra $\mathcal{W}\ S$) → $\mathcal{O} \sqcup \mathcal{V} \sqcup \mathcal{U} \sqcup \mathcal{W}$ ·
$\mathbf{A} \cong \mathbf{B} = \Sigma\ f$ : (hom $\mathbf{A}\ \mathbf{B}$) , $\Sigma\ g$ : (hom $\mathbf{B}\ \mathbf{A}$) , ($\mid f \mid \circ \mid g \mid \sim \mid id\ \mathbf{B} \mid$) × ($\mid g \mid \circ \mid f \mid \sim \mid id\ \mathbf{A} \mid$)

That is, two structures are *isomorphic* provided there are homomorphisms going back and forth between them which compose to the identity.

### 2.3.2 Isomorphism is an equivalence relation

$\cong$-refl : $\{\mathcal{U}$ : Universe$\}$ $\{\mathbf{A}$ : Algebra $\mathcal{U}\ S\}$ → $\mathbf{A} \cong \mathbf{A}$
$\cong$-refl $\{\mathcal{U}\}\{\mathbf{A}\} = id\ \mathbf{A}\ ,\ id\ \mathbf{A}\ ,\ (\lambda\ a \to refl)\ ,\ (\lambda\ a \to refl)$

$\cong$-sym : $\{\mathcal{U}\ \mathcal{W}$ : Universe$\}\{\mathbf{A}$ : Algebra $\mathcal{U}\ S\}\{\mathbf{B}$ : Algebra $\mathcal{W}\ S\}$ → $\mathbf{A} \cong \mathbf{B} \to \mathbf{B} \cong \mathbf{A}$
$\cong$-sym $h =$ fst $\| h \|$ , fst $h$ , $\|$ snd $\| h \| \|$ , $\mid$ snd $\| h \| \mid$

$\cong$-trans : $\{\mathbf{A}$ : Algebra $\mathcal{X}\ S\}\{\mathbf{B}$ : Algebra $\mathcal{Y}\ S\}\{\mathbf{C}$ : Algebra $\mathcal{Z}\ S\}$
            → $\quad \mathbf{A} \cong \mathbf{B} \to \mathbf{B} \cong \mathbf{C} \to \mathbf{A} \cong \mathbf{C}$

$\cong$-trans $\{\mathbf{A}\}\ \{\mathbf{B}\}\{\mathbf{C}\}\ ab\ bc =$ f , g , $\alpha$ , $\beta$
   where
   f1 : hom $\mathbf{A}\ \mathbf{B}$
   f1 $= \mid ab \mid$
   f2 : hom $\mathbf{B}\ \mathbf{C}$
   f2 $= \mid bc \mid$
   f : hom $\mathbf{A}\ \mathbf{C}$
   f $=$ ∘-hom $\mathbf{A}\ \mathbf{C}$ f1 f2

   g1 : hom $\mathbf{C}\ \mathbf{B}$
   g1 $=$ fst $\| bc \|$
   g2 : hom $\mathbf{B}\ \mathbf{A}$
   g2 $=$ fst $\| ab \|$
   g : hom $\mathbf{C}\ \mathbf{A}$
   g $=$ ∘-hom $\mathbf{C}\ \mathbf{A}$ g1 g2

   $\alpha$ : $\mid$ f $\mid \circ \mid$ g $\mid \sim \mid id\ \mathbf{C} \mid$
   $\alpha\ x =$ (ap $\mid$ f2 $\mid$($\mid$ snd $\| ab \| \mid$ ($\mid$ g1 $\mid x$)))·($\mid$ snd $\| bc \| \mid$) $x$

   $\beta$ : $\mid$ g $\mid \circ \mid$ f $\mid \sim \mid id\ \mathbf{A} \mid$
   $\beta\ x =$ (ap $\mid$ g2 $\mid$($\|$ snd $\| bc \| \|$ ($\mid$ f1 $\mid x$)))·($\|$ snd $\| ab \| \|$) $x$

---

[10] For unabridged docs see https://ualib.gitlab.io/Homomorphisms.Isomorphisms.html; for source code see https://gitlab.com/ualib/ualib.gitlab.io/-/blob/master/UALib/Homomorphisms/Isomorphisms.lagda.

To make trans-≅ easier to apply in certain situations, we define a couple of alternatives where the only difference is which arguments are implicit.

```
TRANS-≅ : {A : Algebra 𝒳 S}{B : Algebra 𝒴 S}{C : Algebra 𝒵 S}
  →         A ≅ B → B ≅ C → A ≅ C
TRANS-≅ {A}{B}{C} = trans-≅ A B C

Trans-≅ : (A : Algebra 𝒳 S){B : Algebra 𝒴 S}(C : Algebra 𝒵 S)
  →         A ≅ B → B ≅ C → A ≅ C
Trans-≅ A {B} C = trans-≅ A B C
```

### 2.3.3 Lift is an algebraic invariant

Fortunately, the lift operation preserves isomorphism (i.e., it's an *algebraic invariant*). As algebra is our main focus, this invariance of the lift operation is what makes it a workable solution to the technical problems that arise from the noncumulativity of the universe hierarchy discussed in Prelude.Lifts [3, §2.5].

```
    open Lift

lift-alg-≅ : {A : Algebra 𝒰 S} → A ≅ (lift-alg A 𝒲)
lift-alg-≅ {A} = 𝓁𝒾𝒻𝓉 , 𝓁𝑜𝓌𝑒𝓇 A , extfun lift∼lower , extfun (lower∼lift{𝒲})

lift-alg-hom : (𝒳 : Universe)(𝒴 : Universe){A : Algebra 𝒰 S}(B : Algebra 𝒲 S)
  →            hom A B → hom (lift-alg A 𝒳) (lift-alg B 𝒴)

lift-alg-hom 𝒳 𝒴 {A} B (f , fhom) = lift ∘ f ∘ lower , γ
    where
    lABh : is-homomorphism (lift-alg A 𝒳) B (f ∘ lower)
    lABh = ∘-is-hom (lift-alg A 𝒳) B {lower}{f} (λ _ _ → 𝓇𝑒𝒻𝓁) fhom

    γ : is-homomorphism(lift-alg A 𝒳)(lift-alg B 𝒴) (lift ∘ (f ∘ lower))
    γ = ∘-is-hom (lift-alg A 𝒳) (lift-alg B 𝒴){f ∘ lower}{lift} lABh λ _ _ → 𝓇𝑒𝒻𝓁

lift-alg-iso : {A : Algebra 𝒰 S}{𝒳 : Universe}{B : Algebra 𝒲 S}{𝒴 : Universe}
  →             A ≅ B → (lift-alg A 𝒳) ≅ (lift-alg B 𝒴)

lift-alg-iso A≅B = ≅-trans (≅-trans (≅-sym lift-alg-≅) A≅B) lift-alg-≅
```

### 2.3.4 Lift associativity

The lift is also associative, up to isomorphism at least.

```
lift-alg-assoc : {A : Algebra 𝒰 S} → lift-alg A (𝒲 ⊔ 𝒥) ≅ (lift-alg (lift-alg A 𝒲) 𝒥)
lift-alg-assoc {A} = ≅-trans (≅-trans γ lift-alg-≅) lift-alg-≅
    where
    γ : lift-alg A (𝒲 ⊔ 𝒥) ≅ A
    γ = ≅-sym lift-alg-≅

lift-alg-associative : (A : Algebra 𝒰 S) → lift-alg A (𝒲 ⊔ 𝒥) ≅ (lift-alg (lift-alg A 𝒲) 𝒥)
lift-alg-associative A = lift-alg-assoc {A}
```

### 2.3.5   Products preserve isomorphisms

Products of isomorphic families of algebras are themselves isomorphic. The proof looks a bit technical, but it is as straightforward as it ought to be.

$\prod\cong$ : $\{\mathcal{A} : I \to$ Algebra $\mathcal{U}$ $S\}\{\mathcal{B} : I \to$ Algebra $\mathcal{W}$ $S\} \to \Pi$ $i : I$ , $\mathcal{A}$ $i \cong \mathcal{B}$ $i \to \prod \mathcal{A} \cong \prod \mathcal{B}$

$\prod\cong \{\mathcal{A}\}\{\mathcal{B}\}$ $AB = \gamma$
  where
  $\phi$ : $| \prod \mathcal{A} | \to | \prod \mathcal{B} |$
  $\phi$ $a$ $i =$ | fst $(AB$ $i)$ | $(a$ $i)$

  $\phi$hom : is-homomorphism $(\prod \mathcal{A})$ $(\prod \mathcal{B})$ $\phi$
  $\phi$hom $f$ $a$ = $gfe$ $(\lambda$ $i \to \|$ fst $(AB$ $i)$ $\|$ $f$ $(\lambda$ $x \to a$ $x$ $i))$

  $\psi$ : $| \prod \mathcal{B} | \to | \prod \mathcal{A} |$
  $\psi$ $b$ $i =$ | fst $\|$ $AB$ $i$ $\|$ | $(b$ $i)$

  $\psi$hom : is-homomorphism $(\prod \mathcal{B})$ $(\prod \mathcal{A})$ $\psi$
  $\psi$hom $f$ $\boldsymbol{b}$ = $gfe$ $(\lambda$ $i \to$ snd | snd $(AB$ $i)$ | $f$ $(\lambda$ $x \to \boldsymbol{b}$ $x$ $i))$

  $\phi\sim\psi$ : $\phi \circ \psi \sim$ | $id$ $(\prod \mathcal{B})$ |
  $\phi\sim\psi$ $\boldsymbol{b}$ = $gfe$ $\lambda$ $i \to$ fst $\|$ snd $(AB$ $i)$ $\|$ $(\boldsymbol{b}$ $i)$

  $\psi\sim\phi$ : $\psi \circ \phi \sim$ | $id$ $(\prod \mathcal{A})$ |
  $\psi\sim\phi$ $a$ = $gfe$ $\lambda$ $i \to$ snd $\|$ snd $(AB$ $i)$ $\|$ $(a$ $i)$

  $\gamma$ : $\prod \mathcal{A} \cong \prod \mathcal{B}$
  $\gamma$ = $(\phi$ , $\phi$hom) , $((\psi$ , $\psi$hom) , $\phi\sim\psi$ , $\psi\sim\phi)$

A nearly identical proof goes through for isomorphisms of *lifted* products (though, just for fun, we use the universal quantifier syntax here to express the dependent function type in the statement of the lemma, instead of the Pi notation we used in the statement of the previous lemma; that is, $\forall$ $i \to \mathcal{A}$ $i \cong \mathcal{B}$ (lift $i$) instead of $\Pi$ $i : I$ , $\mathcal{A}$ $i \cong \mathcal{B}$ (lift $i$)).

lift-alg-$\prod\cong$ : $\{I : \mathcal{I}$ `$\}\{\mathcal{A} : I \to$ Algebra $\mathcal{U}$ $S\}\{\mathcal{B} :$ (Lift$\{\mathcal{X}\}$ $I) \to$ Algebra $\mathcal{W}$ $S\}$
  $\to$       $(\forall$ $i \to \mathcal{A}$ $i \cong \mathcal{B}$ (lift $i$)) $\to$ lift-alg $(\prod \mathcal{A})$ $\mathcal{X} \cong \prod \mathcal{B}$

lift-alg-$\prod\cong$ $\{I\}\{\mathcal{A}\}\{\mathcal{B}\}$ $AB = \gamma$
  where
  $\phi$ : $| \prod \mathcal{A} | \to | \prod \mathcal{B} |$
  $\phi$ $a$ $i =$ | fst $(AB$ (lower $i))$ | $(a$ (lower $i))$

  $\phi$hom : is-homomorphism $(\prod \mathcal{A})$ $(\prod \mathcal{B})$ $\phi$
  $\phi$hom $f$ $a$ = $gfe$ $(\lambda$ $i \to (\|$ fst $(AB$ (lower $i))$ $\|)$ $f$ $(\lambda$ $x \to a$ $x$ (lower $i)))$

  $\psi$ : $| \prod \mathcal{B} | \to | \prod \mathcal{A} |$
  $\psi$ $b$ $i =$ | fst $\|$ $AB$ $i$ $\|$ | $(b$ (lift $i))$

  $\psi$hom : is-homomorphism $(\prod \mathcal{B})$ $(\prod \mathcal{A})$ $\psi$
  $\psi$hom $f$ $\boldsymbol{b}$ = $gfe$ $(\lambda$ $i \to$ (snd | snd $(AB$ $i)$ |) $f$ $(\lambda$ $x \to \boldsymbol{b}$ $x$ (lift $i)))$

  $\phi\sim\psi$ : $\phi \circ \psi \sim$ | $id$ $(\prod \mathcal{B})$ |
  $\phi\sim\psi$ $\boldsymbol{b}$ = $gfe$ $\lambda$ $i \to$ fst $\|$ snd $(AB$ (lower $i))$ $\|$ $(\boldsymbol{b}$ $i)$

```
ψ~ϕ : ψ ∘ ϕ ∼ | id (∏ 𝒜) |
ψ~ϕ a = gfe λ i → snd ∥ snd (AB i) ∥ (a i)

A≅B : ∏ 𝒜 ≅ ∏ ℬ
A≅B = (ϕ , ϕhom) , ((ψ , ψhom) , ϕ~ψ , ψ~ϕ)

γ : lift-alg (∏ 𝒜) 𝓧 ≅ ∏ ℬ
γ = ≅-trans (≅-sym lift-alg-≅) A≅B
```

### 2.3.6  Embedding tools

Finally, we prove some useful facts about embeddings that occasionally come in handy.

```
embedding-lift-nat : hfunext 𝓘 𝓤 → hfunext 𝓘 𝓦
  →              {I : 𝓘 ˙}{A : I → 𝓤 ˙}{B : I → 𝓦 ˙}
                 (h : Nat A B) → (∀ i → is-embedding (h i))
                 ————————————————————————————
  →              is-embedding(NatΠ h)

embedding-lift-nat hfu hfw h hem = NatΠ-is-embedding hfu hfw h hem

embedding-lift-nat' : hfunext 𝓘 𝓤 → hfunext 𝓘 𝓦
  →              {I : 𝓘 ˙}{𝒜 : I → Algebra 𝓤 S}{ℬ : I → Algebra 𝓦 S}
                 (h : Nat(fst ∘ 𝒜)(fst ∘ ℬ)) → (∀ i → is-embedding (h i))
                 ————————————————————————————————
  →              is-embedding(NatΠ h)

embedding-lift-nat' hfu hfw h hem = NatΠ-is-embedding hfu hfw h hem

embedding-lift : hfunext 𝓘 𝓤 → hfunext 𝓘 𝓦
  →              {I : 𝓘 ˙} {𝒜 : I → Algebra 𝓤 S}{ℬ : I → Algebra 𝓦 S}
  →              (h : ∀ i → | 𝒜 i | → | ℬ i |) → (∀ i → is-embedding (h i))
                 ————————————————————————————————-
  →              is-embedding(λ (x : | ∏ 𝒜 |) (i : I) → (h i)(x i))

embedding-lift hfu hfw {I}{𝒜}{ℬ} h hem = embedding-lift-nat' hfu hfw {I}{𝒜}{ℬ} h hem

iso→embedding : {𝓤 𝓦 : Universe}{𝐀 : Algebra 𝓤 S}{𝐁 : Algebra 𝓦 S}
  →              (ϕ : 𝐀 ≅ 𝐁) → is-embedding (fst | ϕ |)

iso→embedding ϕ = equivs-are-embeddings (fst | ϕ |) (invertibles-are-equivs (fst | ϕ |) finv)
  where
  finv : invertible (fst | ϕ |)
  finv = | fst ∥ ϕ ∥ | , (snd ∥ snd ϕ ∥ , fst ∥ snd ϕ ∥)
```

## 2.4 Homomorphic Images

This section presents the Homomorphisms.HomomorphicImages module of the AgdaUALib, slightly abridged.[11]

We begin with what seems, for our purposes, the most useful way to represent the class of *homomorphic images* of an algebra in dependent type theory.

HomImage : {**A** : Algebra $\mathcal{U}$ $S$}(**B** : Algebra $\mathcal{W}$ $S$)($\phi$ : hom **A** **B**) → | **B** | → $\mathcal{U}$ ⊔ $\mathcal{W}$ ̇
HomImage **B** $\phi$ = $\lambda$ $b$ → Image | $\phi$ | ∋ $b$

HomImagesOf : Algebra $\mathcal{U}$ $S$ → $\mathcal{O}$ ⊔ $\mathcal{V}$ ⊔ $\mathcal{U}$ ⊔ $\mathcal{W}$ $^+$ ̇
HomImagesOf **A** = Σ **B** : (Algebra $\mathcal{W}$ $S$) , Σ $\phi$ : (| **A** | → | **B** |) , is-homomorphism **A** **B** $\phi$ × Epic $\phi$

These types should be self-explanatory, but just to be sure, let's describe the Sigma type appearing in the second definition. Given an $S$-algebra **A**, the type HomImagesOf **A** denotes the class of algebras **B** : Algebra $\mathcal{W}$ $S$ with a map $\varphi$ : | **A** | → | **B** | such that $\varphi$ is a epimorphism.

The standard (informal) notion of the class of homomorphic images of an algebra assumes closure under isomorphism. Thus, we consider **B** to be a homomorphic image of **A** if (and only if) there exists an algebra **C** which is a homomorphic image of **A** and isomorphic to **B**. In the UALib we express this notion with the following type.

\_is-hom-image-of\_ : (**B** : Algebra $\mathcal{W}$ $S$)(**A** : Algebra $\mathcal{U}$ $S$) → ov $\mathcal{W}$ ⊔ $\mathcal{U}$ ̇
**B** is-hom-image-of **A** = Σ **C**$\phi$ : (HomImagesOf **A**) , | **C**$\phi$ | ≅ **B**

### 2.4.1 Images of a class of algebras

Given a class $\mathcal{K}$ of $S$-algebras, we need a type that expresses the assertion that a given algebra is a *homomorphic image* of some algebra in the class, as well as a type that represents all such homomorphic images.

\_is-hom-image-of-class\_ : Algebra $\mathcal{U}$ $S$ → Pred (Algebra $\mathcal{U}$ $S$)($\mathcal{U}$ $^+$) → ov $\mathcal{U}$ ̇
**B** is-hom-image-of-class $\mathcal{K}$ = Σ **A** : (Algebra $\mathcal{U}$ $S$) , (**A** ∈ $\mathcal{K}$) × (**B** is-hom-image-of **A**)

HomImagesOfClass : Pred (Algebra $\mathcal{U}$ $S$) ($\mathcal{U}$ $^+$) → ov $\mathcal{U}$ ̇
HomImagesOfClass $\mathcal{K}$ = Σ **B** : (Algebra $\mathcal{U}$ $S$) , (**B** is-hom-image-of-class $\mathcal{K}$)

### 2.4.2 Lifting tools

Here are some tools that have been useful (e.g., in the road to the proof of Birkhoff's HSP theorem). The first states and proves the simple fact that the lift of an epimorphism is an epimorphism.

lift-of-alg-epic-is-epic : ($\mathcal{Z}$ : Universe){$\mathcal{W}$ : Universe}
{**A** : Algebra $\mathcal{X}$ $S$}(**B** : Algebra $\mathcal{Y}$ $S$)($h$ : hom **A** **B**)
————————————————
→ Epic | $h$ | → Epic | lift-alg-hom $\mathcal{Z}$ $\mathcal{W}$ **B** $h$ |

---

[11] For unabridged docs see https://ualib.gitlab.io/Homomorphisms.HomomorphicImages.html; for source code see https://gitlab.com/ualib/ualib.gitlab.io/-/blob/master/UALib/Homomorphisms/HomomorphicImages.lagda.

```
lift-of-alg-epic-is-epic 𝓔 {𝓦} {A} B h hepi y = eq y (lift a) η
   where
   lh : hom (lift-alg A 𝓔) (lift-alg B 𝓦)
   lh = lift-alg-hom 𝓔 𝓦 B h

   ζ : Image | h | ∋ (lower y)
   ζ = hepi (lower y)

   a : | A |
   a = Inv | h | ζ

   β : lift (| h | a) ≡ (lift ∘ | h | ∘ lower{𝓦}) (lift a)
   β = ap (λ - → lift (| h | ( - a))) (lower∼lift {𝓦} )

   η : y ≡ | lh | (lift a)
   η = y                ≡⟨ (extfun lift∼lower) y ⟩
       lift (lower y)   ≡⟨ ap lift (InvIsInv | h | ζ)⁻¹ ⟩
       lift (| h | a)   ≡⟨ β ⟩
       | lh | (lift a) ∎

lift-alg-hom-image : {𝓔 𝓦 : Universe}
                    {A : Algebra 𝓧 S}{B : Algebra 𝓨 S}
   →                B is-hom-image-of A
                    ————————————————————————
   →                (lift-alg B 𝓦) is-hom-image-of (lift-alg A 𝓔)

lift-alg-hom-image {𝓔}{𝓦}{A}{B} ((C , φ , φhom , φepic) , C≅B) =
   (lift-alg C 𝓦 , | lφ | , ‖ lφ ‖ , lφepic) , lift-alg-iso C≅B
   where
   lφ : hom (lift-alg A 𝓔) (lift-alg C 𝓦)
   lφ = (lift-alg-hom 𝓔 𝓦 C) (φ , φhom)

   lφepic : Epic | lφ |
   lφepic = lift-of-alg-epic-is-epic 𝓔 C (φ , φhom) φepic
```

## 3    Types for Terms

### 3.1   Basic Definitions

This section presents the Terms.Basic module of the AgdaUALib, slightly abridged.[12] The theoretical background that begins each subsection below is based on Section 4.3 of Cliff Bergman's excellent textbook on universal algebra, [1, §4.3]. Apart from notation, our presentation is similar to Bergman's, but we will try to be concise, omitting some details and examples, in order to more quickly arrive at our objective, which is to use type theory to express the concepts and formalize them in the Agda language. We refer the reader to [1] for a more complete exposition of classical (informal) universal algebra.

---

[12] For unabridged docs see https://ualib.gitlab.io/Terms.Basic.html; for source code see https://gitlab.com/ualib/ualib.gitlab.io/-/blob/master/UALib/Terms/Basic.lagda.

### 3.1.1  The type of terms

Fix a signature $S$ and let $X$ denote an arbitrary nonempty collection of variable symbols. Assume the symbols in $X$ are distinct from the operation symbols of $S$, that is $X \cap \mid S \mid = \emptyset$. By a *word* in the language of $S$, we mean a nonempty, finite sequence of members of $X \cup \mid S \mid$. We denote the concatenation of such sequences by simple juxtaposition. Let $S_0$ denote the set of nullary operation symbols of $S$. We define by induction on $n$ the sets $T_n$ of *words* over $X \cup \mid S \mid$ as follows (cf. [1, Def. 4.19]):

$$T_0 := X \cup S_0 \text{ and } T_{n+1} := T_n \cup \mathscr{T}_n,$$

where $\mathscr{T}_n$ is the collection of all $f\, t$ such that $f : \mid S \mid$ and $t : \parallel S \parallel f \to T_n$. (Recall, $\parallel S \parallel f$ denotes the arity of the operation symbol $f$.)

We define the collection of *terms* in the signature $S$ over $X$ by Term $X := \bigcup_n T_n$. By an *S-term* we mean a term in the language of $S$. Since the definition of Term $X$ is recursive, it would seem that an inductive type could be used to represent the semantic notion of terms in type theory. Indeed, the *inductive type of terms* in the UALib is one such representation; it is defined as follows.

```
data Term {𝒳 : Universe}(X : 𝒳 ˙) : ov 𝒳 ˙ where
  generator : X → Term X
  node : (f : ∣ S ∣)(t : ∥ S ∥ f → Term X) → Term X
```

**Notation**. As usual, the type $X$ represents an arbitrary collection of variable symbols. Recall, ov 𝒳 ˙ is our shorthand notation for the universe 𝒪 ⊔ 𝒱 ⊔ 𝒳 ⁺ ˙. Throughout this module the name of the first constructor of the Term type will remain generator. However, in all of the modules that follow this one, we will use the shorthand $g$ to denote the generator constructor.

### 3.1.2  The term algebra

For a given signature $S$, if the type Term $X$ is nonempty (equivalently, if $X$ or $\mid S \mid$ is nonempty), then we can define an algebraic structure, denoted by **T** $X$, called the *term algebra in the signature S over X*. Since terms take other terms as arguments they do double-duty, serving as both the elements of the domain and the basic operations of the algebra.

- For each operation symbol $f : \mid S \mid$, denote by $f\,\hat{}\,(\mathbf{T}\,X)$ the operation on Term $X$ which maps each tuple $t : \parallel S \parallel f \to \mid \mathbf{T}\,X \mid$ of terms to the formal term $f\, t$.
- Define **T** $X$ to be the algebra with universe $\mid \mathbf{T}\,X \mid := $ Term $X$ and operations $f\,\hat{}\,(\mathbf{T}\,X)$, one for each symbol $f$ in $\mid S \mid$.

  In Agda the term algebra can be defined as simply as one could hope.

  ```
  T : {𝒳 : Universe}(X : 𝒳 ˙) → Algebra (ov 𝒳) S
  T X = Term X , node
  ```

### 3.1.3  The universal property

The term algebra **T** $X$ is *absolutely free* (or *universal* or *initial*) for algebras in the signature $S$. That is, for every $S$-algebra **A**, the following hold.

1. Every function from $X$ to $\mid \mathbf{A} \mid$ lifts to a homomorphism from **T** $X$ to **A**.
2. The homomorphism that exists by item 1 is unique.

We now prove this in Agda, starting with the fact that every map from $X$ to | **A** | lifts to a map from | **T** $X$ | to | **A** | in a natural way, by induction on the structure of a given term.

free-lift : (**A** : Algebra $\mathcal{U}$ $S$)($h$ : $X \to$ | **A** |) $\to$ | **T** $X$ | $\to$ | **A** |
free-lift _ $h$ (generator $x$) = $h$ $x$
free-lift **A** $h$ (node $f$ $t$) = ($f$ ̂ **A**) ($\lambda$ $i \to$ free-lift **A** $h$ ($t$ $i$))

Naturally, at the base step of the induction, when the term has the form generator $x$, the free lift of $h$ agrees with $h$. For the inductive step, when the given term has the form node $f$ $t$, the free lift is defined as follows: Assuming (the induction hypothesis) that we know the image of each subterm $t$ $i$ under the free lift of $h$, define the free lift at the full term by applying $f$ ̂ **A** to the images of the subterms.

The free lift so defined is a homomorphism by construction. Indeed, here is the formal statement and (trivial) proof of this fact.

lift-hom : (**A** : Algebra $\mathcal{U}$ $S$) $\to$ ($X \to$ | **A** |) $\to$ hom (**T** $X$) **A**
lift-hom **A** $h$ = free-lift **A** $h$ , $\lambda$ $f$ $a \to$ ap ($f$ ̂ **A**) $refl$

Finally, we prove that the homomorphism is unique. This requires funext $\mathcal{V}$ $\mathcal{U}$ (i.e., *function extensionality* at universe levels $\mathcal{V}$ and $\mathcal{U}$) which we postulate by making it part of the premise in the following function type definition.

free-unique : funext $\mathcal{V}$ $\mathcal{U}$ $\to$ (**A** : Algebra $\mathcal{U}$ $S$)($g$ $h$ : hom (**T** $X$) **A**)
$\to$         ($\forall$ $x \to$ | $g$ | (generator $x$) $\equiv$ | $h$ | (generator $x$))
              —————————————————————
$\to$         $\forall$ ($t$ : Term $X$) $\to$ | $g$ | $t \equiv$ | $h$ | $t$

free-unique _ _ _ _ $p$ (generator $x$) = $p$ $x$
free-unique $fe$ **A** $g$ $h$ $p$ (node $f$ $t$) = | $g$ | (node $f$ $t$) $\equiv\langle$ ∥ $g$ ∥ $f$ $t$ $\rangle$
                                     ($f$ ̂ **A**)(| $g$ | $\circ$ $t$) $\equiv\langle$ $\alpha$ $\rangle$
                                     ($f$ ̂ **A**)(| $h$ | $\circ$ $t$) $\equiv\langle$ (∥ $h$ ∥ $f$ $t$)$^{-1}$ $\rangle$
                                     | $h$ | (node $f$ $t$) ∎
 where
  $\alpha$ : ($f$ ̂ **A**) (| $g$ | $\circ$ $t$) $\equiv$ ($f$ ̂ **A**) (| $h$ | $\circ$ $t$)
  $\alpha$ = ap ($f$ ̂ **A**) ($fe$ $\lambda$ $i \to$ free-unique $fe$ **A** $g$ $h$ $p$ ($t$ $i$))

Let's account for what we have proved thus far about the term algebra. If we postulate a type $X$ : $\mathfrak{X}$ · (representing an arbitrary collection of variable symbols) such that for each $S$-algebra **A** there is a map from $X$ to the domain of **A**, then it follows that for every $S$-algebra **A** there is a homomorphism from **T** $X$ to | **A** | that "agrees with the original map on $X$," by which we mean that for all $x$ : $X$ the lift evaluated at generator $x$ is equal to the original function evaluated at $x$.

If we further assume that each of the mappings from $X$ to | **A** | is *surjective*, then the homomorphisms constructed with free-lift and lift-hom are *epimorphisms*, as we now prove.

lift-of-epi-is-epi : {**A** : Algebra $\mathcal{U}$ $S$}{$h_0$ : $X \to$ | **A** |}
              ———————————————————-
$\to$              Epic $h_0 \to$ Epic | lift-hom **A** $h_0$ |

lift-of-epi-is-epi {**A**}{$h_0$} $hE$ $y$ = $\gamma$
 where
  $h_0{}^{-1}y$ = Inv $h_0$ ($hE$ $y$)

$$\eta : y \equiv \mid \text{lift-hom } \mathbf{A} \; h_0 \mid (\text{generator } {h_0}^{-1}\text{y})$$
$$\eta = (\text{InvIsInv } h_0 \; (hE \; y))^{-1}$$

$$\gamma : \text{Image} \mid \text{lift-hom } \mathbf{A} \; h_0 \mid \ni y$$
$$\gamma = \text{eq } y \; (\text{generator } {h_0}^{-1}\text{y}) \; \eta$$

The lift-hom and lift-of-epi-is-epi types will be called to action when such epimorphisms are needed later (e.g., in the Varieties module).

## 3.2  Term Operations

This section presents the Terms.Operations module of the AgdaUALib, slightly abridged.[13] Here we define *term operations* which are simply terms interpreted in a particular algebra, and we prove some compatibility properties of term operations.[14]

When we interpret a term in an algebra we call the resulting function a *term operation*. Given a term $p$ and an algebra $\mathbf{A}$, we denote by $p \cdot \mathbf{A}$ the *interpretation* of $p$ in $\mathbf{A}$. This is defined inductively as follows.

1. If $p = x$ (a variable symbol) and $a : X \to \mid \mathbf{A} \mid$ a tuple of elements from the domain of $\mathbf{A}$, then $(p \cdot \mathbf{A}) \; a := a \; x$.
2. If $p = f \; t$ (where $f$ is an operation symbol and $t$ is a tuple of terms) and if $a : X \to \mid \mathbf{A} \mid$ is a tuple from $\mathbf{A}$, then we define $(p \cdot \mathbf{A}) \; a = (f \; t \cdot \mathbf{A}) \; a := (f \; \hat{} \; \mathbf{A}) \; (\lambda \; i \to (t \; i \cdot \mathbf{A}) \; a)$.

Thus the interpretation of a term is defined by induction on the structure of the term, and the definition is formally implemented in the UALib as follows.

    $\_ \cdot \_ : \text{Term } X \to (\mathbf{A} : \text{Algebra } \mathcal{U} \; S) \to (X \to \mid \mathbf{A} \mid) \to \mid \mathbf{A} \mid$
    $(g \; x \cdot \mathbf{A}) \; a = a \; x$
    $(\text{node } f \; t \cdot \mathbf{A}) \; a = (f \; \hat{} \; \mathbf{A}) \; \lambda \; i \to (t \; i \cdot \mathbf{A}) \; a$

It turns out that the interpretation of a term is the same as the free-lift (modulo argument order and assuming function extensionality).

    free-lift-interp : dfunext $\mathcal{V} \; \mathcal{U} \to (\mathbf{A} : \text{Algebra } \mathcal{U} \; S)(h : X \to \mid \mathbf{A} \mid)(p : \text{Term } X)$
    $\to \quad\quad\quad (p \cdot \mathbf{A}) \; h \equiv (\text{free-lift } \mathbf{A} \; h) \; p$
    free-lift-interp $\_ \mathbf{A} \; h \; (g \; x) = \mathit{refl}$
    free-lift-interp $fe \; \mathbf{A} \; h \; (\text{node } f \; t) = \text{ap } (f \; \hat{} \; \mathbf{A}) \; (fe \; \lambda \; i \to \text{free-lift-interp } fe \; \mathbf{A} \; h \; (t \; i))$

If the algebra $\mathbf{A}$ happens to be $\mathbf{T} \; X$, then we expect that $\forall \; s$ we have $(p \cdot \mathbf{T} \; X) \; s \equiv p \; s$. But what is $(p \cdot \mathbf{T} \; X) \; s$ exactly? By definition, it depends on the form of $p$ as follows:

- if $p \equiv g \; x$, then $(p \cdot \mathbf{T} \; X) \; s := (g \; x \cdot \mathbf{T} \; X) \; s \equiv s \; x$;
- if $p = \text{node } f \; t$, then $(p \cdot \mathbf{T} \; X) \; s := (\text{node } f \; t \cdot \mathbf{T} \; X) \; s \equiv (f \; \hat{} \; \mathbf{T} \; X) \; \lambda \; i \to (t \; i \cdot \mathbf{T} \; X) \; s$.

Now, assume $\phi : \text{hom } \mathbf{T} \; \mathbf{A}$. Then by comm-hom-term, we have $\mid \phi \mid (p \cdot \mathbf{T} \; X) \; s \equiv (p \cdot \mathbf{A}) \mid \phi \mid \circ \; s$.

---

[13] For unabridged docs see `https://ualib.gitlab.io/Terms.Operations.html`; for source code see `https://gitlab.com/ualib/ualib.gitlab.io/-/blob/master/UALib/Terms/Operations.lagda`.

[14] **Notation**. At the start of the Terms.Operations module, for notational convenience we rename the generator constructor of the Term type, so that from now on we can use $g$ in place of generator.

- if $p \equiv g\ x$ (and $t : X \to |\ \mathbf{T}\ X\ |$), then

  $|\ \phi\ |\ p := |\ \phi\ |\ (g\ x) \equiv |\ \phi\ |\ (\lambda\ t \to t\ x) \equiv \lambda\ t \to (|\ \phi\ | \circ t)\ x$;
- if $p \equiv \mathsf{node}\ f\ t$, then

  $|\ \phi\ |\ \mathrm{p} := |\ \phi\ |\ (p\ \cdot\ \mathbf{T}\ X)\ s = (\mathsf{node}\ f\ t\ \cdot\ \mathbf{T}\ \mathrm{X})\ s\ (f\ \hat{}\ \mathbf{T}\ X)\ \lambda\ i \to (t\ i\ \cdot\ \mathbf{T}\ X)\ s$.

We claim that for all $p :\ \mathsf{Term}\ X$ there exist $q :\ \mathsf{Term}\ X$ and $t : X \to |\ \mathbf{T}\ X\ |$ such that $p \equiv (q\ \cdot\ \mathbf{T}\ X)\ t$. We prove this fact as follows.

```
term-interp : {𝒳 : Universe}{X : 𝒳 ˙} (f : | S |){s t : ‖ S ‖ f → Term X}
            →           s ≡ t → node f s ≡ (f ^ T X) t
term-interp f {s}{t} st = ap (node f) st

module _ {𝒳 : Universe}{X : 𝒳 ˙}{fe : dfunext 𝒱 (ov 𝒳)} where

  term-gen : (p : | T X |) → Σ q : | T X | , p ≡ (q · T X) g
  term-gen (g x) = (g x) , refl
  term-gen (node f t) = node f(λ i → | term-gen (t i)|) , term-interp f (fe λ i → ‖ term-gen (t i)‖)

  term-gen-agreement : (p : | T X |) → (p · T X) g ≡ (| term-gen p | · T X) g
  term-gen-agreement (g x) = refl
  term-gen-agreement (node f t) = ap (f ^ T X) (fe λ x → term-gen-agreement (t x))

  term-agreement : (p : | T X |) → p ≡ (p · T X) g
  term-agreement p = snd (term-gen p) · (term-gen-agreement p)⁻¹
```

### 3.2.1 Interpretation of terms in product algebras

Note that while in the previous section it sufficed to postulate a local version of function extensionality, in the present section we will assume the full global version (global-dfunext) is in force. (We are not sure whether this is necessary or if, with some effort, we could get a more moderate invocation of function extensionality to work here.)[15]

```
  interp-prod : {𝒲 : Universe}(p : Term X){I : 𝒲 ˙}
                (𝒜 : I → Algebra 𝒰 S)(a : X → ∀ i → | 𝒜 i |)
                ————————————————————————
              →        (p · (∏ 𝒜)) a ≡ (λ i → (p · 𝒜 i) (λ j → a j i))
  interp-prod (g x₁) 𝒜 a = refl

  interp-prod (node f t) 𝒜 a = let IH = λ x → interp-prod (t x) 𝒜 a
                               in
   (f ^ ∏ 𝒜)(λ x → (t x · ∏ 𝒜) a)                    ≡⟨ ap (f ^ ∏ 𝒜)(gfe IH) ⟩
   (f ^ ∏ 𝒜)(λ x → (λ i → (t x · 𝒜 i)(λ j → a j i))) ≡⟨ refl ⟩
   (λ i → (f ^ 𝒜 i) (λ x → (t x · 𝒜 i)(λ j → a j i))) ∎

  interp-prod2 : (p : Term X){I : 𝒰 ˙ }(𝒜 : I → Algebra 𝒰 S)
                 ————————————————————————————
               →        (p · ∏ 𝒜) ≡ λ(t : X → | ∏ 𝒜 |) → (λ i → (p · 𝒜 i)(λ x → t x i))
```

---

[15] We plan to resolve this, and if possible improve upon our treatment of function extensionality, before the next major release of the AgdaUALib.

```
    interp-prod2 (𝑔 x₁) 𝒜 = refl
    interp-prod2 (node f t) 𝒜 = gfe λ (tup : X → | ∏ 𝒜 |) →
      let IH = λ x → interp-prod (t x) 𝒜 in
      let tA = λ z → t z · ∏ 𝒜 in
      (f ̂ ∏ 𝒜)(λ s → tA s tup)                                    ≡⟨ ap(f ̂ ∏ 𝒜)(gfe λ x → IH x tup)⟩
      (f ̂ ∏ 𝒜)(λ s → λ j → (t s · 𝒜 j)(λ ℓ → tup ℓ j))           ≡⟨ refl ⟩
      (λ i → (f ̂ 𝒜 i)(λ s → (t s · 𝒜 i)(λ ℓ → tup ℓ i)))    ∎
```

### 3.2.2 Compatibility of terms

We now prove two important facts about term operations. The first of these, which is used very often in the sequel, asserts that every term commutes with every homomorphism.

```
    comm-hom-term : {𝐀 : Algebra 𝒰 S} (𝐁 : Algebra 𝒲 S)
                    (h : hom 𝐀 𝐁) (t : Term X) (a : X → | 𝐀 |)
                    ─────────────────────────
 →                  | h | ((t · 𝐀) a) ≡ (t · 𝐁) (| h | ∘ a)

    comm-hom-term 𝐁 h (𝑔 x) a = refl
    comm-hom-term {𝐀} 𝐁 h (node f t) a = | h |((f ̂ 𝐀)λ i → (t i · 𝐀) a)    ≡⟨ i ⟩
                                          (f ̂ 𝐁)(λ i → | h |((t i · 𝐀) a)) ≡⟨ ii ⟩
                                          (f ̂ 𝐁)(λ r → (t r · 𝐁)(| h | ∘ a)) ∎
      where
      i = ∥ h ∥ f(λ r → (t r · 𝐀) a)
      ii = ap (f ̂ 𝐁)(gfe (λ i → comm-hom-term 𝐁 h (t i) a))
```

To conclude the Terms module, we prove that every term is compatible with every congruence relation. That is, if t : Term X and θ : Con 𝐀, then a θ b → t(a) θ t(b).

```
    open Congruence
    compatible-term : {𝐀 : Algebra 𝒰 S}(t : Term X)(θ : Con 𝐀)
                      ─────────────────────────
 →                    compatible-fun (t · 𝐀) | θ |

    compatible-term (𝑔 x) θ p = p x
    compatible-term (node f t) θ p = snd ∥ θ ∥ f λ x → (compatible-term (t x) θ) p
```

## 4 Subalgebra Types

### 4.1 Subuniverses

This section presents the Subalgebras.Subuniverses module of the AgdaUALib, slightly abridged.[16] We start by defining a type that represents the important concept of *subuniverse*. Suppose 𝐀 is an algebra. A subset B ⊆ | 𝐀 | is said to be *closed under the operations of* 𝐀 if for each f ∈ | S | and all tuples b : ∥ S ∥ f → B the element (f ̂ 𝐀) b belongs to B. If a subset B ⊆ A is closed under the operations of 𝐀, then we call B a *subuniverse* of 𝐀.

---

[16] For unabridged docs see https://ualib.gitlab.io/Subalgebras.Subuniverses.html; for source code see https://gitlab.com/ualib/ualib.gitlab.io/-/blob/master/UALib/Subalgebras/Subuniverses.lagda.

We first show how to represent in Agda the collection of subuniverses of an algebra **A**. Since a subuniverse is viewed as a subset of the domain of **A**, we define it as a predicate on | **A** |. Thus, the collection of subuniverses is a predicate on predicates on | **A** |.

Subuniverses : (**A** : Algebra $\mathcal{U}$ $S$) → Pred (Pred | **A** | $\mathcal{W}$)($\mathcal{O}$ ⊔ $\mathcal{V}$ ⊔ $\mathcal{U}$ ⊔ $\mathcal{W}$)
Subuniverses **A** $B$ = ($f$ : | $S$ |)($a$ : ‖ $S$ ‖ $f$ → | **A** |) → Im $a$ ⊆ $B$ → ($f$ ̂ **A**) $a$ ∈ $B$

An algebra can be constructed out of a subuniverse in the following natural way.

SubunivAlg : (**A** : Algebra $\mathcal{U}$ $S$)($B$ : Pred | **A** | $\mathcal{W}$) → $B$ ∈ Subuniverses **A** → Algebra ($\mathcal{U}$ ⊔ $\mathcal{W}$) $S$
SubunivAlg **A** $B$ $B{\in}SubA$ = Σ $B$ , λ $f$ $b$ → ($f$ ̂ **A**)(fst ∘ $b$) , $B{\in}SubA$ $f$ (fst ∘ $b$)(snd ∘ $b$)

### 4.1.1 Subuniverses as records

Next we define a type to represent a single subuniverse of an algebra. If **A** is the algebra in question, then the subuniverse will be a subset of (i.e., predicate over) the domain '| **A** |' that belongs to Subuniverses **A**.

record Subuniverse {**A** : Algebra $\mathcal{U}$ $S$} : ov ($\mathcal{U}$ ⊔ $\mathcal{W}$) ˙ where
  constructor mksub
  field
    sset : Pred | **A** | $\mathcal{W}$
    isSub : sset ∈ Subuniverses **A**

As an example application, here is a formal proof that the equalizer of two homomorphisms with domain **A** is a subuniverse of **A**.

$E$hom-is-subuniverse : dfunext $\mathcal{V}$ $\mathcal{W}$ → {**A** : Algebra $\mathcal{U}$ $S$}(**B** : Algebra $\mathcal{W}$ $S$)($g$ $h$ : hom **A** **B**)
    →                   Subuniverse {**A** = **A**}

$E$hom-is-subuniverse $fe$ **B** $g$ $h$ =
    mksub ($E$hom{$fe$ = $fe$} **B** $g$ $h$) λ $f$ $a$ $x$ → $E$hom-closed{$fe$ = $fe$} **B** $g$ $h$ $f$ $a$ $x$

### 4.1.2 Subuniverse Generation

If **A** is an algebra and $X$ ⊆ | **A** | a subset of the domain of **A**, then the *subuniverse of* **A** *generated by* $X$ is typically denoted by Sg$^{\mathbf{A}}$ ($X$) and defined to be the smallest subuniverse of **A** containing $X$. Equivalently,

Sg$^{\mathbf{A}}$ ($X$) = ⋂ { $U$ : $U$ is a subuniverse of **A** and $B$ ⊆ $U$}.

We define an inductive type, denoted by Sg, that represents the subuniverse generated by a given subset of the domain of a given algebra, as follows.

data Sg (**A** : Algebra $\mathcal{U}$ $S$)($X$ : Pred | **A** | $\mathcal{W}$) : Pred | **A** | ($\mathcal{O}$ ⊔ $\mathcal{V}$ ⊔ $\mathcal{W}$ ⊔ $\mathcal{U}$) where
  var : ∀ {$v$} → $v$ ∈ $X$ → $v$ ∈ Sg **A** $X$
  app : ($f$ : | $S$ |)($a$ : ‖ $S$ ‖ $f$ → | **A** |) → Im $a$ ⊆ Sg **A** $X$ → ($f$ ̂ **A**) $a$ ∈ Sg **A** $X$

Given an arbitrary $S$-algebra **A** and subset $X$ of | **A** |, the type Sg $X$ does indeed represent a subuniverse of **A**; proving this with the inductive type Sg is trivial, as we see here.

sgIsSub : {**A** : Algebra $\mathcal{U}$ $S$}{$X$ : Pred | **A** | $\mathcal{W}$} → Sg **A** $X$ ∈ Subuniverses **A**
sgIsSub = app

Next we prove by structural induction that Sg $X$ is the smallest subuniverse of **A** containing $X$.

   sgIsSmallest : {𝓡 : Universe}(**A** : Algebra 𝓤 S){X : Pred | **A** | 𝓦}(Y : Pred | **A** | 𝓡)
    →            Y ∈ Subuniverses **A** → X ⊆ Y → Sg **A** X ⊆ Y

   sgIsSmallest _ _ _ XinY (var Xv) = XinY Xv

   sgIsSmallest **A** Y YsubA XinY (app f a SgXa) = fa∈Y
     where
     IH : Im a ⊆ Y
     IH i = sgIsSmallest **A** Y YsubA XinY (SgXa i)

     fa∈Y : (f ̂ **A**) a ∈ Y
     fa∈Y = YsubA f a IH

When the inhabitant of Sg X is constructed as app f a SgXa, we may assume (the induction hypothesis) that the arguments in the tuple a belong to Y. Then the result of applying f to a also belongs to Y since Y is a subuniverse.

### 4.1.3 Subuniverse Lemmas

Here we formalize a few basic properties of subuniverses. First, the intersection of subuniverses is again a subuniverse.

   sub-intersection : {**A** : Algebra 𝓤 S}{I : 𝓘 ˙}{𝒜 : I → Pred | **A** | 𝓦}
    →                 (Π i : I , 𝒜 i ∈ Subuniverses **A**) → ⋂ I 𝒜 ∈ Subuniverses **A**

   sub-intersection α f a β = λ i → α i f a λ j → β j i

In the proof above, we assume the following typing judgments:
α :   ∀ i → 𝒜 i ∈ Subuniverses **A**
f :   | S |
a :   ‖ S ‖ f → | **A** |
β :   Im a ⊆ ⋂ I 𝒜
and we must prove (f ̂ **A**) a ∈ ⋂ I 𝒜. In this case, Agda will fill in the proof term λ i → α i f a (λ x → β x i) automatically with the command C-c C-a in agda2-mode.

Next we formalize the proof that subuniverses are closed under the action of term operations.

   sub-term-closed : {𝒳 : Universe}{X : 𝒳 ˙}(**A** : Algebra 𝓤 S){B : Pred | **A** | 𝓦}
    →                (B ∈ Subuniverses **A**) → (t : Term X)(b : X → | **A** |)
    →                (∀ x → b x ∈ B) → ((t · **A**) b) ∈ B
   sub-term-closed **A** α (g x) b Bb = Bb x
   sub-term-closed **A** {B} α (node f t) b β =
     α f (λ z → (t z · **A**) b) λ x → sub-term-closed **A** {B} α (t x) b β

In the induction step of the foregoing proof, the typing judgments of the premise are these:
**A** :   Algebra 𝓤 S    B :   Pred | **A** | 𝓦       α :   B ∈ Subuniverses **A**
f :   | S |              t :   ‖ S ‖ f → Term X
b :   X → | **A** |      β :   ∀ x → b x ∈ B

This is another instance in which Agda will fill in the correct proof term if we invoke the command C-c C-a in agda2-mode.

Alternatively, we could express the preceding fact using an inductive type representing images of terms.

```
data TermImage (A : Algebra 𝒰 S)(Y : Pred | A | 𝒲) : Pred | A | (𝒪 ⊔ 𝒱 ⊔ 𝒰 ⊔ 𝒲)
  where
  var : ∀ {y : | A |} → y ∈ Y → y ∈ TermImage A Y
  app : ∀ f t → Π x : ∥ S ∥ f , t x ∈ TermImage A Y → (f ̂ A) t ∈ TermImage A Y
```

By what we proved above, it should come as no surprise that TermImage A Y is a subuniverse of A that contains Y. Indeed, the proof is trivial.

```
TermImageIsSub : {A : Algebra 𝒰 S}{Y : Pred | A | 𝒲} → TermImage A Y ∈ Subuniverses A
TermImageIsSub = app

Y-onlyif-TermImageY : {A : Algebra 𝒰 S}{Y : Pred | A | 𝒲} → Y ⊆ TermImage A Y
Y-onlyif-TermImageY {a} Ya = var Ya
```

Since Sg A Y is the smallest subuniverse containing Y, we obtain the following inclusion.

```
SgY-onlyif-TermImageY : (A : Algebra 𝒰 S)(Y : Pred | A | 𝒲) → Sg A Y ⊆ TermImage A Y
SgY-onlyif-TermImageY A Y = sgIsSmallest A(TermImage A Y) TermImageIsSub Y-onlyif-TermImageY
```

### 4.1.4   Homomorphic images are subuniverses

Now that we have developed the machinery of subuniverse generation, we can prove two basic facts that play an important role in many theorems about algebraic structures. First, the image of a homomorphism is a subuniverse of its codomain.

```
hom-image-is-sub : {A : Algebra 𝒰 S}{B : Algebra 𝒲 S}
                   (φ : hom A B) → (HomImage B φ) ∈ Subuniverses B

hom-image-is-sub {A}{B} φ f b Imfb = eq ((f ̂ B) b) ((f ̂ A) ar) γ
  where
  ar : ∥ S ∥ f → | A |
  ar = λ x → Inv | φ | (Imfb x)

  ζ : | φ | ∘ ar ≡ b
  ζ = gfe (λ x → InvIsInv | φ | (Imfb x))

  γ : (f ̂ B) b ≡ | φ | ((f ̂ A) ar)
  γ = (f ̂ B) b             ≡⟨ ap (f ̂ B)(ζ ⁻¹) ⟩
      (f ̂ B) (| φ | ∘ ar)  ≡⟨(∥ φ ∥ f ar)⁻¹ ⟩
      | φ | ((f ̂ A) ar)   ∎
```

Next we prove the important fact that homomorphisms are uniquely determined by their values on a generating set.

```
hom-unique : funext 𝒱 𝒲 → {A : Algebra 𝒰 S}{B : Algebra 𝒲 S}
             (X : Pred | A | 𝒰) (g h : hom A B)
  →          Π x : | A | , (x ∈ X → | g | x ≡ | h | x)
             ———————————————————————-
  →          Π a : | A | , (a ∈ Sg A X → | g | a ≡ | h | a)

hom-unique _ _ _ _ α a (var x) = α a x

hom-unique fe {A}{B} X g h α fa (app f 𝒂 β) = | g | ((f ̂ A) 𝒂)  ≡⟨ ∥ g ∥ f 𝒂 ⟩
```

$$(f \mathbin{\hat{}} \mathbf{B})(|\, g\, |\, \circ\, \boldsymbol{a}\,) \equiv \langle\ \mathsf{ap}\ (f \mathbin{\hat{}} \mathbf{B})(fe\ \mathsf{IH})\ \rangle$$
$$(f \mathbin{\hat{}} \mathbf{B})(|\, h\, |\, \circ\, \boldsymbol{a}) \equiv \langle\ (\ \|\, h\, \|\, f\, \boldsymbol{a}\,)^{-1}\ \rangle$$
$$|\, h\, |\, ((f \mathbin{\hat{}} \mathbf{A})\, \boldsymbol{a}\,)\ \blacksquare$$

where $\mathsf{IH} = \lambda\, x \to$ hom-unique $fe\ \{\mathbf{A}\}\{\mathbf{B}\}\ X\ g\ h\ \alpha\ (\boldsymbol{a}\ x)\ (\beta\ x)$

In the induction step of the foregoing proof, the typing judgments of the premise are these:

| | | | | | |
|---|---|---|---|---|---|
| $fe$ : | funext $\mathcal{V}\ \mathcal{W}$ | $\mathbf{A}$ : | Algebra $\mathcal{U}\ S$ | $\mathbf{B}$ : | Algebra $\mathcal{W}\ S$ |
| $X$ : | Pred $\|\,\mathbf{A}\,\|\,\mathcal{U}$ | $g\ h$ : | hom $\mathbf{A}\ \mathbf{B}$ | $\alpha$ : | $\Pi\ x : |\,\mathbf{A}\,|\,,\ (x \in X \to |\,g\,|\,x \equiv |\,h\,|\,x)$ |
| $fa$ : | $|\,\mathbf{A}\,|$ | $f$ : | $|\,S\,|$ | $a$ : | $\|\,S\,\|\,f \to |\,\mathbf{A}\,|$ |
| $\beta$ : | Im $a \subseteq$ Sg $\mathbf{A}\ X$ | | | | |

where $fa = (f \mathbin{\hat{}} \mathbf{A})\ a$. Under these assumptions, we prove $|\,g\,|\,((f \mathbin{\hat{}} \mathbf{A})\ a) \equiv |\,h\,|\,((f \mathbin{\hat{}} \mathbf{A})\ a)$.

## 4.2 Subalgebras

This section presents the Subalgebras.Subalgebras module of the AgdaUALib, slightly abridged.[17] Here we define the Subalgebra type, representing the subalgebra of a given algebra, as well as the collection of all subalgebras of a given class of algebras.

### 4.2.1 Subalgebra type

Given algebras $\mathbf{A}$ : Algebra $\mathcal{U}\ S$ and $\mathbf{B}$ : Algebra $\mathcal{W}\ S$, we say that $\mathbf{B}$ is a *subalgebra* of $\mathbf{A}$ just in case $\mathbf{B}$ can be *homomorphically embedded* in $\mathbf{A}$; i.e., there exists a map $h : |\,\mathbf{B}\,| \to |\,\mathbf{A}\,|$ that is both a homomorphism and an embedding.[18]

    \_IsSubalgebraOf\_ : $\{\mathcal{W}\ \mathcal{U}\ :\ \mathsf{Universe}\}(\mathbf{B}\ :\ \mathsf{Algebra}\ \mathcal{W}\ S)(\mathbf{A}\ :\ \mathsf{Algebra}\ \mathcal{U}\ S) \to \mathcal{O} \sqcup \mathcal{V} \sqcup \mathcal{U} \sqcup \mathcal{W}$ ˙
    $\mathbf{B}$ IsSubalgebraOf $\mathbf{A} = \Sigma\ h : \mathsf{hom}\ \mathbf{B}\ \mathbf{A}\ ,\ \mathsf{is\text{-}embedding}\ |\,h\,|$

    Subalgebra : $\{\mathcal{W}\ \mathcal{U}\ :\ \mathsf{Universe}\} \to \mathsf{Algebra}\ \mathcal{U}\ S \to \mathsf{ov}\ \mathcal{W} \sqcup \mathcal{U}$ ˙
    Subalgebra $\{\mathcal{W}\}\ \mathbf{A} = \Sigma\ \mathbf{B} : (\mathsf{Algebra}\ \mathcal{W}\ S)\ ,\ \mathbf{B}\ \mathsf{IsSubalgebraOf}\ \mathbf{A}$

Note the order of the arguments. The universe $\mathcal{W}$ comes first because in certain situations we have to explicitly specify this universe, whereas we can almost always leave the universe $\mathcal{U}$ implicit. See, for example, the definition of IsSubalgebraOfClass below (§4.2.3).

### 4.2.2 Consequences of First Homomorphism Theorem

We take this opportunity to prove an important lemma that makes use of the IsSubalgebraOf type defined above; it is the following: If $\mathbf{A}$ and $\mathbf{B}$ are $S$-algebras and $h$ : hom $\mathbf{A}\ \mathbf{B}$ a homomorphism from $\mathbf{A}$ to $\mathbf{B}$, then the quotient $\mathbf{A}\ /\ \mathsf{ker}\ h$ is (isomorphic to) a subalgebra of $\mathbf{B}$. This is an easy corollary of the First Homomorphism Theorem proved in the Homomorphisms.Noether module.

    FirstHomCorollary : $\{\mathcal{U}\ \mathcal{W}\ :\ \mathsf{Universe}\}$
                           – *extensionality assumptions* –

---

[17] For unabridged docs see https://ualib.gitlab.io/Subalgebras.Subalgebras.html; for source code see https://gitlab.com/ualib/ualib.gitlab.io/-/blob/master/UALib/Subalgebras/Subalgebras.lagda.

[18] An alternative which could end up being simpler and easier to work with would be to proclaim $\mathbf{B}$ a subalgebra of $\mathbf{A}$ iff there is an *injective* homomorphism from $B$ into $\mathbf{A}$. In preparation for the next major release of the UALib, we will investigate the consequences of taking that path instead of the stricter embedding requirement we chose for the definition of the type IsSubalgebraOf.

$\rightarrow$                         dfunext $\mathcal{V}$ $\mathcal{W}$ $\rightarrow$ prop-ext $\mathcal{U}$ $\mathcal{W}$

$\rightarrow$                         (**A** : Algebra $\mathcal{U}$ $S$)(**B** : Algebra $\mathcal{W}$ $S$)($h$ : hom **A** **B**)

                                – *truncation assumptions* –
$\rightarrow$                         is-set | **B** |
$\rightarrow$                         ($\forall$ $a$ $x$ $\rightarrow$ is-subsingleton ($\langle$ kercon **B** $h$ $\rangle$ $a$ $x$))
$\rightarrow$                         ($\forall$ $C$ $\rightarrow$ is-subsingleton ($\mathcal{C}\{A =$ | **A** |$\}\{\langle$ kercon **B** $h$ $\rangle\}$ $C$))
                                ————————————————————————
$\rightarrow$                         (**A** [ **B** ]/ker $h$) IsSubalgebraOf **B**

FirstHomCorollary $fe$ $pe$ **A** **B** $h$ $Bset$ $ssR$ $ssA$ = $\phi$hom , $\phi$emb
 where
 FirstHomThm : $\Sigma$ $\phi$ : hom (**A** [ **B** ]/ker $h$) **B** , (| $h$ | $\equiv$ | $\phi$ | $\circ$ | $\pi$ker **B** $h$ | )
                                                               $\times$ Monic | $\phi$ | $\times$ is-embedding | $\phi$ |
 FirstHomThm = FirstHomomorphismTheorem $fe$ $pe$ **A** **B** $h$ $Bset$ $ssR$ $ssA$

 $\phi$hom : hom (**A** [ **B** ]/ker $h$) **B**
 $\phi$hom = | FirstHomThm |

 $\phi$emb : is-embedding | $\phi$hom |
 $\phi$emb = snd (snd (snd FirstHomThm))

One special case to which we will apply this is where the algebra **A** is the term algebra **T** $X$. We formalize this special case here so that it's readily available when we need it later.

 free-quot-subalg : $\{\mathcal{U}$ $\mathcal{X}$ : Universe$\}$
                         –*extensionality assumptions* –
$\rightarrow$                         dfunext $\mathcal{V}$ $\mathcal{U}$ $\rightarrow$ prop-ext (ov $\mathcal{X}$) $\mathcal{U}$

$\rightarrow$                         ($X$ : $\mathcal{X}$ $\cdot$)(**B** : Algebra $\mathcal{U}$ $S$)($h$ : hom (**T** $X$) **B**)

                                –*truncation assumptions* –
$\rightarrow$                         is-set | **B** |
$\rightarrow$                         ($\forall$ $p$ $q$ $\rightarrow$ is-subsingleton ($\langle$ kercon **B** $h$ $\rangle$ $p$ $q$))
$\rightarrow$                         ($\forall$ $C$ $\rightarrow$ is-subsingleton ($\mathcal{C}\{A =$ | **T** $X$ |$\}\{\langle$ kercon **B** $h$ $\rangle\}$ $C$))
                                ————————————————————————
$\rightarrow$                         ((**T** $X$) [ **B** ]/ker $h$) IsSubalgebraOf **B**

 free-quot-subalg $fe$ $pe$ $X$ **B** $h$ $Bset$ $ssR$ $ssB$ = FirstHomCorollary $fe$ $pe$ (**T** $X$) **B** $h$ $Bset$ $ssR$ $ssB$

**Notation.** For convenience, we define the following shorthand for the subalgebra relation.

 _$\leq$_ : $\{\mathcal{W}$ $\mathcal{U}$ : Universe$\}$(**B** : Algebra $\mathcal{W}$ $S$)(**A** : Algebra $\mathcal{U}$ $S$) $\rightarrow$ $\mathcal{O}$ $\sqcup$ $\mathcal{V}$ $\sqcup$ $\mathcal{U}$ $\sqcup$ $\mathcal{W}$ $\cdot$
 **B** $\leq$ **A** = **B** IsSubalgebraOf **A**

From now on we will use **B** $\leq$ **A** to express the assertion that **B** is a subalgebra of **A**.

### 4.2.3 Subalgebras of a class

One of our goals is to formally express and prove properties of classes of algebraic structures. Fixing a signature $S$ and a universe $\mathcal{U}$, we define a class of $S$-algebras with domains of type $\mathcal{U}$ $\cdot$ as a predicate over the Algebra $\mathcal{U}$ $S$ type. In the syntax of the UALib, such predicates inhabit the type Pred (Algebra $\mathcal{U}$ $S$) $\mathcal{Z}$, for some universe $\mathcal{Z}$.

Suppose 𝒦 : Pred (Algebra 𝒰 S) 𝒵 denotes a class of S-algebras and **B** : Algebra 𝒲 S denotes an arbitrary S-algebra. Then we might wish to consider the assertion that **B** is a subalgebra of some algebra in the class 𝒦. The next type we define allows us to express this assertion as **B** IsSubalgebraOfClass 𝒦.

```
_IsSubalgebraOfClass_ : Algebra 𝒲 S → Pred (Algebra 𝒰 S) 𝒵 → ov (𝒰 ⊔ 𝒲) ⊔ 𝒵 ˙
B IsSubalgebraOfClass 𝒦 = Σ A : Algebra 𝒰 S , Σ sa : Subalgebra{𝒲} A , (A ∈ 𝒦) × (B ≅ ∣ sa ∣)
```

Using this type, we express the collection of all subalgebras of algebras in a class by the type SubalgebraOfClass, which we now define.

```
SubalgebraOfClass : {𝒲 𝒰 : Universe} → Pred (Algebra 𝒰 S)(ov 𝒰) → ov (𝒰 ⊔ 𝒲) ˙
SubalgebraOfClass {𝒲} 𝒦 = Σ B : Algebra 𝒲 S , B IsSubalgebraOfClass 𝒦
```

### 4.2.4 Subalgebra lemmas

We conclude this module by proving a number of useful facts about subalgebras. Some of the formal statements below may appear to be redundant, and indeed they are to some extent. However, each one differs slightly from the next, if only with respect to the explicitness or implicitness of their arguments. The aim is to make it as convenient as possible to apply the lemmas in different situations. (We're stocking the UALib utility closet now; elegance is not the priority.)

First we show that the subalgebra relation is a *preorder*. Recall, this means it is reflexive and transitive.[19]

```
≤-reflexive : {𝒰 : Universe}(A : Algebra 𝒰 S) → A ≤ A
≤-reflexive A = (id ∣ A ∣ , id-is-hom) , id-is-embedding

≤-refl : {𝒰 : Universe}{A : Algebra 𝒰 S} → A ≤ A
≤-refl {𝒰}{A} = ≤-reflexive A

≤-transitivity : (A : Algebra 𝒳 S)(B : Algebra 𝒴 S)(C : Algebra 𝒵 S)
 →          C ≤ B → B ≤ A → C ≤ A
≤-transitivity A B C CB BA = (∘-hom C A ∣ CB ∣ ∣ BA ∣) , ∘-embedding ∥ BA ∥ ∥ CB ∥

≤-trans : (A : Algebra 𝒳 S){B : Algebra 𝒴 S}{C : Algebra 𝒵 S}
 →      C ≤ B → B ≤ A → C ≤ A
≤-trans A {B}{C} = ≤-transitivity A B C
```

Next we prove that if two algebras are isomorphic and one of them is a subalgebra of **A**, then so is the other.

```
≤-iso : (A : Algebra 𝒳 S){B : Algebra 𝒴 S}{C : Algebra 𝒵 S}
 →    C ≅ B → B ≤ A → C ≤ A
≤-iso A {B} {C} CB BA = (g ∘ f , gfhom) , gfemb
   where
```

---

[19] In the Relations.Quotients module, we defined *preorder* for binary relation types. Here, however, we will content ourselves with merely proving reflexivity and transitivity of the subalgebra relation ≤, without worry about first defining it as an inhabitant of an honest-to-goodness binary relation type, of the sort introduced in the Relations.Discrete module. Perhaps we will address this matter in a future release of the UALib.

```
f : | C | → | B |
f = fst | CB |
g : | B | → | A |
g = fst | BA |

gfemb : is-embedding (g ∘ f)
gfemb = ∘-embedding (‖ BA ‖) (iso→embedding CB)

gfhom : is-homomorphism C A (g ∘ f)
gfhom = ∘-is-hom C A {f}{g} (snd | CB |) (snd | BA |)
```

The following variations on this theme are sometimes useful.

```
≤-trans-≅ : (A : Algebra 𝒳 S){B : Algebra 𝒴 S}(C : Algebra 𝒵 S)
     →            A ≤ B → A ≅ C → C ≤ B
≤-trans-≅ A {B} C A≤B B≅C = ≤-iso B (≅-sym B≅C) A≤B

≤-TRANS-≅ : (A : Algebra 𝒳 S){B : Algebra 𝒴 S}(C : Algebra 𝒵 S)
     →            A ≤ B → B ≅ C → A ≤ C
≤-TRANS-≅ A C A≤B B≅C = ( ∘-hom A C | A≤B | | B≅C |) ,
                              ∘-embedding (iso→embedding B≅C)(‖ A≤B ‖)
```

Next we prove a monotonicity property of ≤.

```
≤-mono : {𝒲 𝒰 𝒵 : Universe}(B : Algebra 𝒲 S){𝒦 𝒦' : Pred (Algebra 𝒰 S) 𝒵}
     →       𝒦 ⊆ 𝒦' → B IsSubalgebraOfClass 𝒦 → B IsSubalgebraOfClass 𝒦'
≤-mono B KK' KB = | KB | , fst ‖ KB ‖ , KK'(| snd ‖ KB ‖ |) , ‖ (snd ‖ KB ‖) ‖
```

Later we will require a number of facts having to do with the relationship between the subalgebra relation and the lifting of algebras to higher universe levels. Here are the tools we need.

```
lift-alg-is-sub : {𝒰 : Universe}{𝒦 : Pred (Algebra 𝒰 S)(ov 𝒰)}{B : Algebra 𝒰 S}
     →           B IsSubalgebraOfClass 𝒦 → (lift-alg B 𝒰) IsSubalgebraOfClass 𝒦
lift-alg-is-sub (A , (sa , (KA , B≅sa))) = A , sa , KA , ≅-trans (≅-sym lift-alg-≅) B≅sa

lift-alg-≤ : (A : Algebra 𝒳 S){B : Algebra 𝒴 S} → B ≤ A → lift-alg B 𝒵 ≤ A
lift-alg-≤ A {B} B≤A = ≤-iso A (≅-sym lift-alg-≅) B≤A

≤-lift-alg : (A : Algebra 𝒳 S){B : Algebra 𝒴 S} → B ≤ A → B ≤ lift-alg A 𝒵
≤-lift-alg A {B} B≤A = ≤-TRANS-≅ B {A} (lift-alg A 𝒵) B≤A lift-alg-≅

lift-alg-≤-lift : {A : Algebra 𝒳 S}(B : Algebra 𝒴 S) → A ≤ B → lift-alg A 𝒵 ≤ lift-alg B 𝒲

lift-alg-≤-lift {A} B A≤B = ≤-trans (lift-alg B 𝒲) (≤-trans B lAA A≤B) B≤lB
    where

    lAA : (lift-alg A 𝒵) ≤ A
    lAA = lift-alg-≤ A {A} ≤-refl

    B≤lB : B ≤ lift-alg B 𝒲
    B≤lB = ≤-lift-alg B {B} ≤-refl
```

## 5 Concluding Remarks

We've reached the end of the second installment in our three-part series describing the AgdaUALib. The next part [4] covers free algebras, equational classes of algebras (i.e., varieties), and Birkhoff's HSP theorem.

We conclude by noting that one of our goals is to make computer formalization of mathematics more accessible to mathematicians working in universal algebra and model theory. We welcome feedback from the community and are happy to field questions about the UALib, how it is installed, and how it can be used to prove theorems that are not yet part of the library. Merge requests submitted to the UALib's main gitlab repository are especially welcomed. Please visit the repository at https://gitlab.com/ualib/ualib.gitlab.io/ and help us improve it.

### References

1. Clifford Bergman. *Universal Algebra: fundamentals and selected topics*, volume 301 of *Pure and Applied Mathematics (Boca Raton)*. CRC Press, Boca Raton, FL, 2012.
2. Ana Bove and Peter Dybjer. *Dependent Types at Work*, pages 57–99. Springer Berlin Heidelberg, Berlin, Heidelberg, 2009. doi:10.1007/978-3-642-03153-3_2.
3. William DeMeo. The Agda Universal Algebra Library, Part 1: Foundation. *CoRR*, abs/2103.05581, 2021. Source code available at https://gitlab.com/ualib/ualib.gitlab.io. URL: https://arxiv.org/abs/2101.10166, arXiv:2103.05581.
4. William DeMeo. The Agda Universal Algebra Library, Part 3: Identity. *CoRR*, 2021. (to appear). URL: http://arxiv.org/a/demeo_w_1.
5. Martín Hötzel Escardó. Introduction to univalent foundations of mathematics with Agda. *CoRR*, abs/1911.00580, 2019. URL: http://arxiv.org/abs/1911.00580, arXiv:1911.00580.
6. Emmanuel Gunther, Alejandro Gadea, and Miguel Pagano. Formalization of universal algebra in Agda. *Electronic Notes in Theoretical Computer Science*, 338:147 – 166, 2018. The 12th Workshop on Logical and Semantic Frameworks, with Applications (LSFA 2017). URL: http://www.sciencedirect.com/science/article/pii/S1571066118300768, doi:https://doi.org/10.1016/j.entcs.2018.10.010.
7. Ulf Norell. Dependently typed programming in agda. In *International school on advanced functional programming*, pages 230–266. Springer, 2008.
8. Ulf Norell. Dependently typed programming in Agda. In *Proceedings of the 6th International Conference on Advanced Functional Programming*, AFP'08, pages 230–266, Berlin, Heidelberg, 2009. Springer-Verlag. URL: http://dl.acm.org/citation.cfm?id=1813347.1813352.
9. The Univalent Foundations Program. *Homotopy Type Theory: Univalent Foundations of Mathematics*. Lulu and The Univalent Foundations Program, Institute for Advanced Study, 2013. URL: https://homotopytypetheory.org/book.
10. The Agda Team. Agda Language Reference section on Axiom K, 2021. URL: https://agda.readthedocs.io/en/v2.6.1/language/without-k.html.
11. The Agda Team. Agda Language Reference section on Safe Agda, 2021. URL: https://agda.readthedocs.io/en/v2.6.1/language/safe-agda.html#safe-agda.
12. The Agda Team. Agda Tools Documentation section on Pattern matching and equality, 2021. URL: https://agda.readthedocs.io/en/v2.6.1/tools/command-line-options.html#pattern-matching-and-equality.
13. Philip Wadler, Wen Kokke, and Jeremy G. Siek. *Programming Language Foundations in Agda*. July 2020. URL: http://plfa.inf.ed.ac.uk/20.07/.